%% file: main.tex
\documentclass[11pt]{article}

\ifdefined\pdfminorversion
\fi

\input{preamble}


\begin{document}

\thispagestyle{empty}
\noindent
\includegraphics[width=3.3cm]{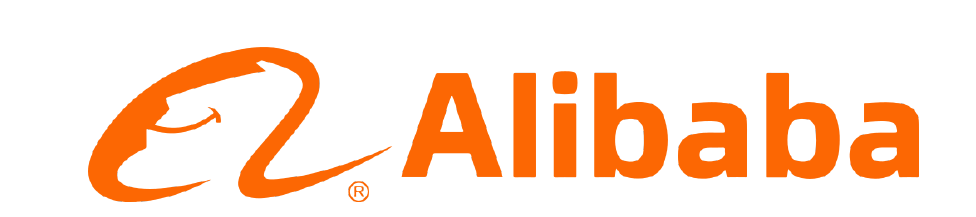}
\hfill
{\itshape August 2026}
\par\vspace{0.35em}\hrule\vspace{1.5em}

\begin{center}
  {\fontsize{25}{29}\selectfont\bfseries RecGPT-Mobile-V2 Technical Report\par}
  \vspace{0.9em}
  {\large\bfseries RecGPT-Mobile Team\par}
\end{center}

\vspace{1.0em}
\input{sections/00_abstract}

\vspace{0.35em}
\begin{figure}[H]
    \centering
    \includegraphics[width=0.90\textwidth]{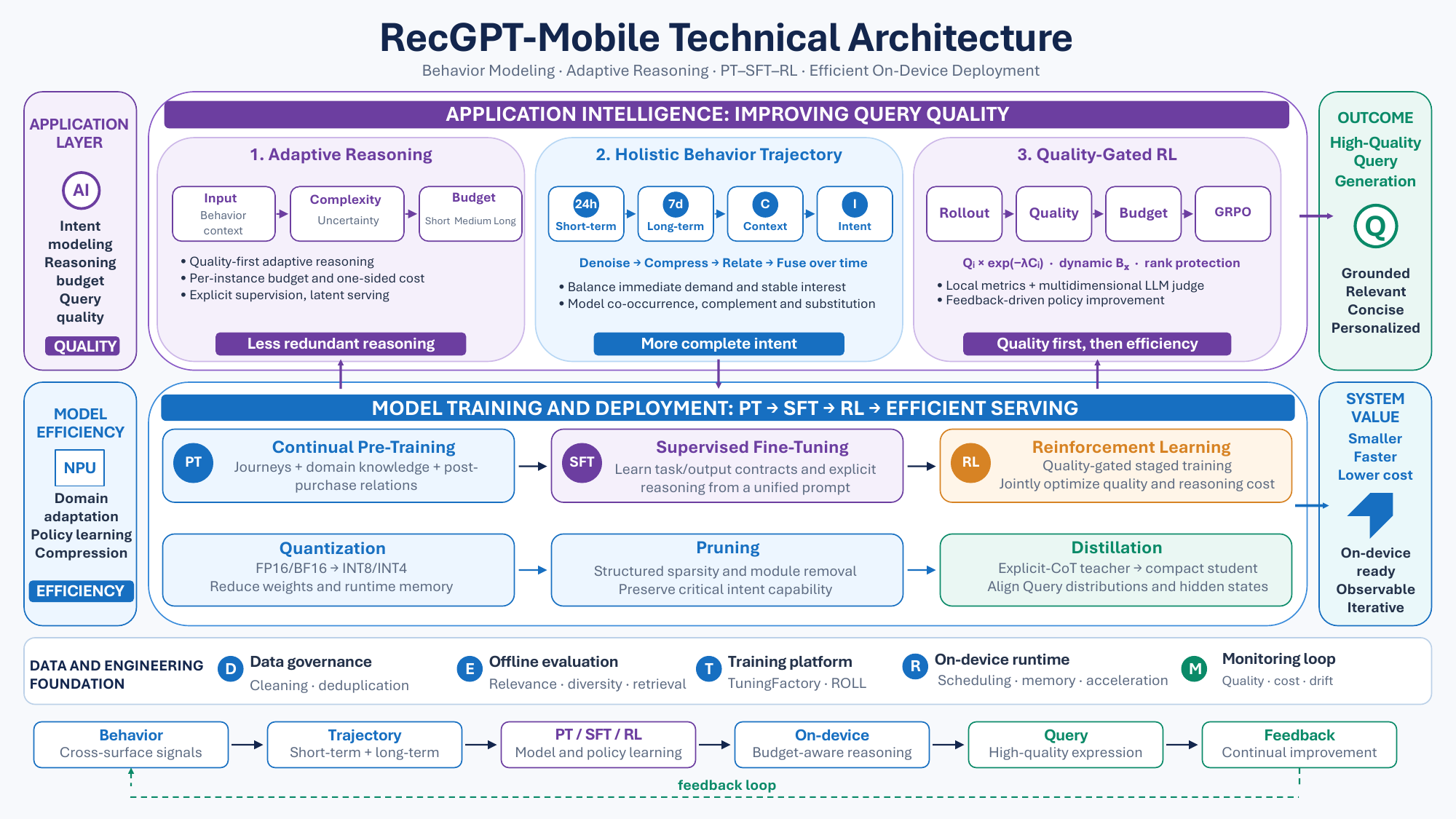}
    \caption{\modelname architecture. Application-level intelligence improves Query quality and compute allocation, while staged training and compression enable practical on-device deployment.}
    \label{fig:roadmap}
\end{figure}

\vfill
\hrule\vspace{0.45em}
{\small\textcopyright\ 2026 Alibaba. All rights reserved.}
\clearpage

\section*{\contentsname}
\begingroup
\footnotesize
\setlength{\parskip}{0pt}
\input{sections/contents_static}
\endgroup
\clearpage

\input{sections/01_introduction}
\input{sections/02_related_work}
\input{sections/03_problem_formulation}
\input{sections/04_framework}
\input{sections/05_continual_pretraining}
\input{sections/06_supervised_finetuning}
\input{sections/07_quality_gated_rl}
\input{sections/08_compression_deployment}
\input{sections/09_experimental_setup}
\input{sections/10_results_analysis}
\input{sections/12_conclusion}

{\small\input{bibliography_static}}

\appendix
\input{sections/appendix_contributors}
\input{sections/appendix_prompt_examples}

\end{document}

%% file: preamble.tex
\usepackage[a4paper,top=2.05cm,bottom=2.15cm,left=2.2cm,right=2.2cm]{geometry}
\usepackage{XCharter}
\usepackage{amsmath,amssymb}
\usepackage{booktabs,tabularx,array}
\usepackage{graphicx}
\usepackage{caption}
\usepackage{enumitem}
\usepackage{xcolor}
\usepackage{natbib}
\usepackage{hyperref}
\usepackage{fancyhdr}
\usepackage{fvextra}
\usepackage{float}
\usepackage[section]{placeins}
\usepackage{setspace}
\usepackage{titlesec}
\usepackage{xspace}
\graphicspath{{figures/}}

\definecolor{recnavy}{HTML}{13264D}
\definecolor{recpurple}{HTML}{6C3EA0}
\definecolor{recblue}{HTML}{176FC1}
\definecolor{recgreen}{HTML}{0A8A68}
\definecolor{reclight}{HTML}{F3F6FA}
\definecolor{recmuted}{HTML}{5D6B82}
\definecolor{recslate}{HTML}{66899A}
\definecolor{recpromptblue}{HTML}{DDEEF8}
\definecolor{recpromptgold}{HTML}{FFF1D9}
\definecolor{reccore}{HTML}{7A1C16}

\hypersetup{
  colorlinks=true,
  linkcolor=recnavy,
  citecolor=recpurple,
  urlcolor=recblue,
  pdfauthor={Alibaba RecGPT-Mobile Team},
  pdftitle={RecGPT-Mobile Technical Report}
}

\titleformat{\section}{\Large\bfseries}{\thesection.}{0.65em}{}
\titleformat{\subsection}{\large\bfseries}{\thesubsection.}{0.65em}{}
\titleformat{\subsubsection}{\normalsize\bfseries}{\thesubsubsection.}{0.65em}{}
\titlespacing*{\section}{0pt}{1.7ex plus .3ex minus .2ex}{.8ex}
\titlespacing*{\subsection}{0pt}{1.35ex plus .2ex minus .1ex}{.5ex}

\setlist[itemize]{leftmargin=1.6em,itemsep=.25em,topsep=.35em}
\setlist[enumerate]{leftmargin=1.8em,itemsep=.3em,topsep=.35em}
\DeclareCaptionLabelSeparator{reportpipe}{\enspace|\enspace}
\newcolumntype{Y}{>{\raggedright\arraybackslash}X}
\newcolumntype{P}[1]{>{\raggedright\arraybackslash}p{#1}}
\newcommand{\modelname}{\textsc{RecGPT-Mobile}\xspace}

\newcommand{\promptbar}[2]{%
  \noindent\colorbox{#1}{%
    \parbox{\dimexpr\linewidth-2\fboxsep\relax}{\color{white}\bfseries #2}%
  }\par\smallskip
}
\newenvironment{systemprompt}[1][System]
  {\par\medskip\promptbar{recslate}{#1}\begingroup}
  {\endgroup\par\medskip}
\newenvironment{userprompt}[1][User]
  {\par\medskip\promptbar{recblue}{#1}\begingroup}
  {\endgroup\par\medskip}

\newcommand{\promptinput}[1]{%
  \VerbatimInput[
    fontsize=\footnotesize,
    breaklines=true,
    breakanywhere=true,
    breaksymbolleft={},
    breaksymbolright={}
  ]{#1}%
}

\DeclareMathOperator{\Quantile}{Quantile}
\DeclareMathOperator{\clip}{clip}

%% file: sections/00_abstract.tex
\begingroup
\small\bfseries
\sloppy
\noindent
Personalized Query prediction maps implicit behavioral signals---clicks, favorites, purchases, and post-purchase exploration---to explicit retrieval intent. On-device deployment makes this task particularly challenging: behavioral trajectories are noisy and multi-scale, multiple Queries may be valid for a single trajectory, and a uniform reasoning policy either expends unnecessary computation on simple instances or allocates insufficient capacity to complex ones. We introduce \textsc{RecGPT-Mobile-V2}, an end-to-end framework that treats intent quality and execution efficiency as coupled objectives within a staged design. The framework transforms heterogeneous interactions into an evidence-preserving trajectory, establishes a recommendation-native foundation through domain adaptation and supervised alignment, and applies reasoning-cost optimization only after grouped rollouts meet grounding and utility criteria. The resulting teacher is distilled into a compact student deployed with low-bit execution, structured compression, and budget-aware device--cloud routing. In an aligned CoT ablation, an evidence-focused short rationale increases ROUGE-L from $0.228$ to $0.315$ and Jaccard from $0.174$ to $0.248$, while slightly outperforming the full five-stage rationale. In the controlled RL comparison, the complete reward formulation improves Query quality from $73.2\%$ under quality-only RL to $78.6\%$, lowers the hard-failure rate from $3.6\%$ to $1.6\%$, and reduces the median CoT length from $62$ to $14$ tokens. Online retrieval analysis further indicates that the Query recall channel retrieves inventory complementary to that surfaced by established recall channels. Collectively, these findings support sufficiency-oriented rather than uniformly short reasoning: retain decision-relevant evidence and allocate additional computation only when it is likely to improve the predicted Query.
\par
\endgroup

\vspace{0.55em}
\noindent\small\textbf{Keywords:} personalized Query prediction; on-device language model; behavioral trajectory; adaptive reasoning; reinforcement learning; model compression

%% file: sections/contents_static.tex
\contentsline {section}{\numberline {1}Introduction}{4}{section.1}%
\contentsline {section}{\numberline {2}Related Work}{5}{section.2}%
\contentsline {subsection}{\numberline {2.1}Intent-Centric Recommendation and Query Generation}{5}{subsection.2.1}%
\contentsline {subsection}{\numberline {2.2}Semantic IDs and Recommendation-Native Language Models}{5}{subsection.2.2}%
\contentsline {subsection}{\numberline {2.3}From Long Reasoning to Adaptive Computation}{5}{subsection.2.3}%
\contentsline {subsection}{\numberline {2.4}Latent Reasoning and Mobile Execution}{5}{subsection.2.4}%
\contentsline {section}{\numberline {3}Problem Setting}{6}{section.3}%
\contentsline {subsection}{\numberline {3.1}Input Representation}{6}{subsection.3.1}%
\contentsline {subsection}{\numberline {3.2}Teacher and Student Outputs}{6}{subsection.3.2}%
\contentsline {subsection}{\numberline {3.3}Hard Failures}{6}{subsection.3.3}%
\contentsline {subsection}{\numberline {3.4}Quality--Efficiency Objective}{6}{subsection.3.4}%
\contentsline {section}{\numberline {4}System Overview}{7}{section.4}%
\contentsline {subsection}{\numberline {4.1}Behavior-Trajectory Compression}{7}{subsection.4.1}%
\contentsline {subsection}{\numberline {4.2}Offline Learning Pipeline}{8}{subsection.4.2}%
\contentsline {subsection}{\numberline {4.3}Online Serving Pipeline}{9}{subsection.4.3}%
\contentsline {subsection}{\numberline {4.4}Design Principles}{9}{subsection.4.4}%
\contentsline {section}{\numberline {5}Learning a Recommendation-Native Foundation}{9}{section.5}%
\contentsline {subsection}{\numberline {5.1}Compact Hybrid Backbone and Selective Adaptation}{9}{subsection.5.1}%
\contentsline {subsection}{\numberline {5.2}General-Capability Recovery}{9}{subsection.5.2}%
\contentsline {subsection}{\numberline {5.3}Recommendation-Domain Continual Pre-Training}{10}{subsection.5.3}%
\contentsline {subsection}{\numberline {5.4}Hierarchical Semantic Item Representation}{10}{subsection.5.4}%
\contentsline {subsection}{\numberline {5.5}Validation and Stage Boundary}{11}{subsection.5.5}%
\contentsline {section}{\numberline {6}Task Alignment with Structured Reasoning}{11}{section.6}%
\contentsline {subsection}{\numberline {6.1}Unified Prompt and Output Contract}{11}{subsection.6.1}%
\contentsline {subsection}{\numberline {6.2}Five-Stage Teacher Rationale}{12}{subsection.6.2}%
\contentsline {subsection}{\numberline {6.3}Evidence-First Short Reasoning}{12}{subsection.6.3}%
\contentsline {subsection}{\numberline {6.4}Aligned CoT Information Ablation}{12}{subsection.6.4}%
\contentsline {subsection}{\numberline {6.5}Initialization and Checkpoint Selection}{13}{subsection.6.5}%
\contentsline {section}{\numberline {7}Quality-Gated Adaptive Reasoning}{13}{section.7}%
\contentsline {subsection}{\numberline {7.1}Why a Naive Length Penalty Fails}{14}{subsection.7.1}%
\contentsline {subsection}{\numberline {7.2}Grouped Rollouts and Quality Qualification}{14}{subsection.7.2}%
\contentsline {subsection}{\numberline {7.3}Evaluator Governance and Calibration}{14}{subsection.7.3}%
\contentsline {subsection}{\numberline {7.4}Input-Specific Budget and One-Sided Cost}{15}{subsection.7.4}%
\contentsline {subsection}{\numberline {7.5}Multiplicative Reward and Rank Protection}{15}{subsection.7.5}%
\contentsline {subsection}{\numberline {7.6}Grouped Policy Update}{15}{subsection.7.6}%
\contentsline {subsection}{\numberline {7.7}Mechanism Ablations}{16}{subsection.7.7}%
\contentsline {subsection}{\numberline {7.8}Target Policy Behavior}{16}{subsection.7.8}%
\contentsline {section}{\numberline {8}On-Device Mixed-Precision Quantization for Inference}{16}{section.8}%
\contentsline {subsection}{\numberline {8.1}Quantization Space and Tensor Granularity}{16}{subsection.8.1}%
\contentsline {subsection}{\numberline {8.2}Sensitivity-Aware Bit Allocation}{17}{subsection.8.2}%
\contentsline {subsection}{\numberline {8.3}Calibration and Outlier Protection}{17}{subsection.8.3}%
\contentsline {subsection}{\numberline {8.4}Weight, Activation, and Cache Precision}{17}{subsection.8.4}%
\contentsline {subsection}{\numberline {8.5}Quantization-Aware Recovery}{17}{subsection.8.5}%
\contentsline {subsection}{\numberline {8.6}Hardware and Quality Validation}{18}{subsection.8.6}%
\contentsline {section}{\numberline {9}Experimental Design and Metrics}{18}{section.9}%
\contentsline {subsection}{\numberline {9.1}Experiment Matrix}{18}{subsection.9.1}%
\contentsline {subsection}{\numberline {9.2}Behavior-Compression Evaluation}{18}{subsection.9.2}%
\contentsline {subsection}{\numberline {9.3}Query Quality Metrics}{18}{subsection.9.3}%
\contentsline {subsection}{\numberline {9.4}Domain PT and SFT Initialization Diagnostic}{19}{subsection.9.4}%
\contentsline {subsection}{\numberline {9.5}CoT Information and Stage Ablations}{19}{subsection.9.5}%
\contentsline {subsection}{\numberline {9.6}Retrieval Complementarity}{19}{subsection.9.6}%
\contentsline {subsection}{\numberline {9.7}Mechanism Validation and Common-Horizon Comparison}{20}{subsection.9.7}%
\contentsline {subsection}{\numberline {9.8}RL Quality--Cost Metrics}{20}{subsection.9.8}%
\contentsline {subsection}{\numberline {9.9}Device Metrics}{20}{subsection.9.9}%
\contentsline {subsection}{\numberline {9.10}Statistical and Acceptance Protocol}{21}{subsection.9.10}%
\contentsline {section}{\numberline {10}Results and Analysis}{21}{section.10}%
\contentsline {subsection}{\numberline {10.1}Behavior Compression Produces a Stable Context Envelope}{21}{subsection.10.1}%
\contentsline {subsection}{\numberline {10.2}Domain Initialization Establishes an Optimization--Retention Trade-off}{21}{subsection.10.2}%
\contentsline {subsection}{\numberline {10.3}What CoT Adds: Evidence and Selective Benefit}{22}{subsection.10.3}%
\contentsline {subsection}{\numberline {10.4}The Query Path Retrieves Complementary Inventory}{23}{subsection.10.4}%
\contentsline {subsection}{\numberline {10.5}RL Training Dynamics Validate the Reward Mechanisms}{24}{subsection.10.5}%
\contentsline {subsection}{\numberline {10.6}Controlled Common-Horizon Comparison}{25}{subsection.10.6}%
\contentsline {subsection}{\numberline {10.7}Structured Quality Explains the A6 Gain}{26}{subsection.10.7}%
\contentsline {subsection}{\numberline {10.8}Serving Consequence of the Shorter Policy}{27}{subsection.10.8}%
\contentsline {subsection}{\numberline {10.9}Combined Interpretation}{27}{subsection.10.9}%
\contentsline {section}{\numberline {11}Conclusion}{28}{section.11}%
\contentsline {section}{\numberline {A}Contributors}{31}{appendix.A}%
\contentsline {section}{\numberline {B}Prompt Examples}{32}{appendix.B}%
\contentsline {subsection}{\numberline {B.1}Reasoning-Need Annotation and Structured Summary}{32}{subsection.B.1}%
\contentsline {subsection}{\numberline {B.2}Post-Purchase Complementarity Judge}{34}{subsection.B.2}%
\contentsline {subsection}{\numberline {B.3}Explicit-CoT Query Generation}{35}{subsection.B.3}%

%% file: sections/01_introduction.tex
\section{Introduction}

Modern commerce systems observe a rich stream of implicit signals across recommendation, search, content, and transaction surfaces. These signals reveal what a user viewed, compared, saved, or purchased, but they do not directly reveal what the user would type next. Personalized Query prediction bridges this gap. Its objective is to infer the retrieval intent most likely to emerge from recent behavior and express that intent as a concise, grounded, and searchable Query.

Post-purchase behavior makes the problem more than a sequence-to-text mapping. A consumable purchase may imply replenishment; an apparel purchase may trigger a search for a matching item; an electronic device may lead to accessories, supplies, or support services. The model must distinguish evidence from incidental activity, connect events across categories and time, decide which interpretation is actionable, and phrase the result at an appropriate level of specificity. In other words, Query prediction requires a small but meaningful reasoning process: \emph{evidence attribution, intent formation, and Query expression}.

Long chain-of-thought models can make this process explicit, but their default behavior is poorly matched to mobile serving. Straightforward repurchase cases and ambiguous cross-category cases often receive similar reasoning depth. Repetition, unnecessary branching, and self-confirmation increase decoding time, cache pressure, energy consumption, and tail latency without necessarily changing the final Query. At the other extreme, forcing every request into a short answer can remove the evidence combinations that difficult intents require. Efficient reasoning is therefore not synonymous with uniformly short reasoning. It requires assigning the minimum sufficient computation to each request.

The quality constraint is also unusual. In mathematics, a response can often be verified as correct or incorrect. Future-Query prediction inherently admits multiple valid answers and is only partially supervised: several phrasings may express the same intent, historical labels reflect previous exposure, and a semantically plausible Query may still be too broad, too narrow, or unsupported by behavior. A length reward built on a weak quality signal will simply make the model exploit that signal more efficiently. \modelname therefore defines reasoning efficiency under a Query-quality constraint rather than as an independent brevity target.

On-device deployment adds a second layer to the problem. Keeping behavioral context local improves privacy and responsiveness, yet a compact model cannot reproduce the verbose textual reasoning of a cloud-scale teacher on every request. Our approach separates the form of reasoning used for training from the form used for serving. Explicit rationales remain available to supervise and audit the teacher. The mobile student learns to retain the useful intermediate information in compact internal representations and exposes only the final Query. A lightweight mechanism based on complexity and confidence decides how much local computation to spend and when a request should be escalated.

This report presents the resulting system as one connected path from behavior to Query. Its main contributions are:

\begin{enumerate}
    \item \textbf{Evidence-preserving behavioral modeling.} We combine fine-grained recent events, compressed longer-term history, current context, and stable profile features while retaining the temporal and item-relation cues needed to separate immediate demand from persistent preference.
    \item \textbf{A recommendation-native training progression.} Domain-adaptive pre-training teaches the model item semantics and behavioral relations; supervised fine-tuning establishes the Query contract and an auditable reasoning structure; reinforcement learning then improves quality and removes unnecessary reasoning.
    \item \textbf{Quality-gated adaptive reasoning.} For multiple rollouts of the same input, reasoning length is optimized only among quality-qualified candidates. A per-input budget, one-sided excess cost, and quality-preserving reward shaping prevent short but poor Queries from being treated as efficient.
    \item \textbf{A deployment-oriented teacher--student design.} Quantization, structured compression, distillation, compact thinking representations, and device--cloud routing translate the trained capability into a practical mobile execution path.
    \item \textbf{Task-specific empirical evidence.} Controlled reasoning ablations show that evidence-focused short rationales provide the strongest quality--cost trade-off, while retrieval-overlap analysis indicates that the Query-prediction channel retrieves inventory complementary to that of established recall channels.
\end{enumerate}

The remainder of the report follows this path. We first summarize the relevant literature and formulate the task. We then describe the system from trajectory construction to domain learning, task alignment, adaptive reasoning, and deployment. Finally, we organize the empirical evidence around the decisions it supports and discuss the limits of offline evaluation.

%% file: sections/02_related_work.tex
\section{Related Work}

\subsection{Intent-Centric Recommendation and Query Generation}

Sequential recommenders such as SASRec and BERT4Rec model preference evolution from ordered interactions, while DIN emphasizes target-aware selection of behavior relevant to a candidate item \citep{kang2018sasrec,sun2019bert4rec,zhou2018din}. These systems predict items within a catalog. Personalized Query prediction instead produces an open-ended textual action, so the model must preserve collaborative and temporal evidence while learning lexical specificity, behavioral grounding, and retrieval utility.

Generative Query research has progressively moved from typed-prefix completion to context-conditioned intent prediction. Behavioral-hypothesis models aggregate earlier Queries and clicks to generate a later Query; more recent work applies preference optimization or unified generative architectures to improve relevance and business utility \citep{chen2020behavioralquery,ouyang2025tppo,guo2025onesug}. The present setting is distinguished by heterogeneous cross-surface behavior and post-purchase inference: the evidence may contain no explicit prefix, and the useful next Query can target a complementary item, a replenishment need, a substitute, or a newly formed intent.

This report also builds on the RecGPT lineage. RecGPT reframes industrial recommendation around explicit user-intent modeling, and RecGPT-V2 develops hierarchical reasoning, compressed behavior representations, constrained reinforcement learning, and structured model-based evaluation \citep{yi2025recgpt,yi2025recgptv2}. The SIGIR~'26 RecGPT-Mobile system establishes on-device next-Query prediction from recent Taobao behavior \citep{zhang2026recgptmobile}. Our focus is complementary: we provide a mechanism-level account of behavior compression, recommendation-domain initialization, CoT information value, adaptive reasoning control, evaluator governance, and the quality--cost evidence needed for a mobile deployment path.

\subsection{Semantic IDs and Recommendation-Native Language Models}

Generative retrieval replaces independent item classification with autoregressive generation of semantic identifiers. TIGER demonstrates that hierarchical codewords can preserve semantic neighborhoods and improve generalization to sparsely observed items \citep{rajput2023tiger}. This motivates the joint item interface used here: semantic IDs retain collaborative structure, while titles, categories, and selected attributes keep the codes grounded in human-readable product semantics. The interface is learned before task alignment so that SFT and RL can operate on recommendation-native representations rather than requiring a small model to infer the domain entirely from the final Query labels.

\subsection{From Long Reasoning to Adaptive Computation}

Long CoT enables decomposition, alternative-path exploration, and self-correction, but applying these behaviors uniformly incurs unnecessary computation on easy inputs \citep{chen2025reasoningera,yeo2025demystifying}. Existing efficiency methods intervene at three levels. Model-level approaches train models for concise or variable-length reasoning through reinforcement learning and supervised tuning \citep{lou2025adacot,kang2024c3ot,xia2025tokenskip,ma2025cotvalve}. Input-level approaches estimate difficulty or specify a token budget before generation \citep{han2024tale,ong2024routellm}. Output-level approaches compress intermediate computation instead of verbalizing every step.

The optimization layer in this report uses Group Relative Policy Optimization, introduced by DeepSeekMath as a critic-free, group-relative alternative to PPO \citep{shao2024deepseekmath}. Length-control studies such as L1 and DAST further show that the useful budget is input-dependent and that the length preference must be conditioned on correctness \citep{aggarwal2025l1,shen2025dast}. Query prediction requires a richer qualification rule than exact-answer verification can provide, so our grouped objective combines label consistency, behavioral support, structured task judgment, and hard output checks before comparing reasoning cost.

\subsection{Latent Reasoning and Mobile Execution}

Continuous and compressed reasoning methods show that useful intermediate computation need not remain in natural-language form. Coconut feeds hidden states back as continuous thoughts, CODI transfers explicit reasoning into continuous representations, and LightThinker summarizes completed reasoning into a compact state \citep{hao2024coconut,shen2025codi,zhang2025lightthinker}. This direction is particularly suitable for recommendation and search because the intermediate rationale is an internal computation rather than a user-facing explanation.

Mobile deployment nevertheless requires more than output-token reduction. Weight storage, activation traffic, and runtime variance must also be controlled, and device capability limits must be respected. The complete design therefore combines evidence-aware trajectory compression, a recommendation-native backbone, adaptive reasoning, quantization and structured compression, teacher--student transfer, and device--cloud routing. The complete system is evaluated as a single progression from behavioral evidence to a deployable Query policy rather than as a collection of independent compression techniques.

%% file: sections/03_problem_formulation.tex
\section{Problem Setting}
\label{sec:problem}

\subsection{Input Representation}

At prediction time, the model receives
\begin{equation}
    x = (H_{\mathrm{recent}}, H_{\mathrm{long}}, c, p),
    \label{eq:input}
\end{equation}
where $H_{\mathrm{recent}}$ is a fine-grained recent behavior sequence, $H_{\mathrm{long}}$ is a compressed longer-term journey, $c$ represents the current interaction context, and $p$ contains stable profile features permitted by the data policy. An event is represented as
\begin{equation}
    e_j=(a_j,v_j,s_j,\Delta t_j,m_j),
\end{equation}
where $a_j$ is the action, $v_j$ is an item or content representation, $s_j$ is the source surface, $\Delta t_j$ is relative time, and $m_j$ contains approved task-relevant metadata. Active signals such as searching, favoriting, adding to a cart, and purchasing are not treated as equivalent in evidential value to passive exposure or interface navigation.

Product identity combines readable text with hierarchical semantic item tokens. Coarse semantic codes locate an item in a broad collaborative neighborhood; finer codes distinguish nearby products. The text prevents the identifiers from becoming ungrounded symbols, while the identifiers retain collaborative information that titles alone may miss. The tokenizer, semantic-code mapping, behavior template, and Query normalizer are versioned together so that pre-training, SFT, RL, and evaluation share the same interpretation of the input.

\subsection{Teacher and Student Outputs}

During teacher training, an output is
\begin{equation}
    y_{\mathrm{teacher}}=(z,q),
\end{equation}
where $z$ is an explicit rationale and $q$ is a single final Query. The rationale is placed in its own delimited region and is never included when final-Query metrics are computed. Reasoning length $L(z)$ counts only rationale tokens; prompt tokens, boundary markers, and Query tokens are excluded.

The deployed student emits no visible CoT. It produces a compact internal sequence $u_{1:K_x}$ and decodes only the Query:
\begin{equation}
    q\sim p_{\theta_s}(q\mid x,u_{1:K_x}).
\end{equation}
The request-dependent computation, governed by $K_x$, can collapse to a direct path for obvious intents or expand for cases that require evidence comparison. This definition keeps the teacher's reasoning observable while the student remains efficient.

\subsection{Hard Failures}

Certain outputs are invalid regardless of their similarity score. They include an empty or unparsable Query, multiple Queries in the final region, explanations after the final Query, forced truncation before a legal ending, and unsupported critical entities. These checks are deterministic and override the continuous score. Keeping them outside the learned evaluator prevents a favorable semantic judgment from compensating for obvious protocol failures.

\subsection{Quality--Efficiency Objective}

The complete objective maximizes expected Query quality while satisfying distinct reasoning and deployment constraints:
\pagebreak
\begin{align}
    \max_{\theta}\quad &\mathbb{E}_{x}[Q(x,q_{\theta})],\\
    \text{s.t.}\quad
    &\mathbb{E}_{x}[C_{\mathrm{reason}}(x;\theta)]\leq \mathcal{B}_{r},\\
    &\operatorname{P95Latency}(\theta,d)\leq \mathcal{B}_{l}(d),\\
    &\operatorname{Memory}(\theta,d)\leq \mathcal{B}_{m}(d),
\end{align}
where $d$ denotes a device tier. Reasoning tokens, device latency, and peak memory are kept separate because they respond to different interventions: adaptive CoT changes the amount of decoded computation, quantization changes memory traffic, pruning changes the executed structure, and routing changes where the request is served.

The formulation exposes the central asymmetry of the task. Quality is mandatory; brevity is valuable only after quality is secured. The training and deployment mechanisms in the following sections are organized around that ordering.

%% file: sections/04_framework.tex
\section{System Overview}
\label{sec:framework}

The first-page overview in Figure~1 summarizes the system as two connected layers. The upper layer asks \emph{what the model should infer}: it constructs a holistic trajectory, allocates reasoning according to ambiguity, and learns from quality-aware feedback. The lower layer asks \emph{how that capability should be trained and served}: it progresses from domain learning to task alignment and policy optimization, then applies compression and teacher--student transfer for device execution.

\subsection{Behavior-Trajectory Compression}
\label{sec:behavior-compression}

The input pipeline does not send the raw event stream directly to the model. It applies the five-layer transformation in Figure~2 and Table~1. The first two layers remove noise and repeated actions; the third restores meaning to identifiers; the fourth exposes evidence strength; and the final layer produces a compact sequence that can be consumed consistently by all learning stages.

\begin{figure}[!htbp]
    \centering
    \includegraphics[width=\textwidth]{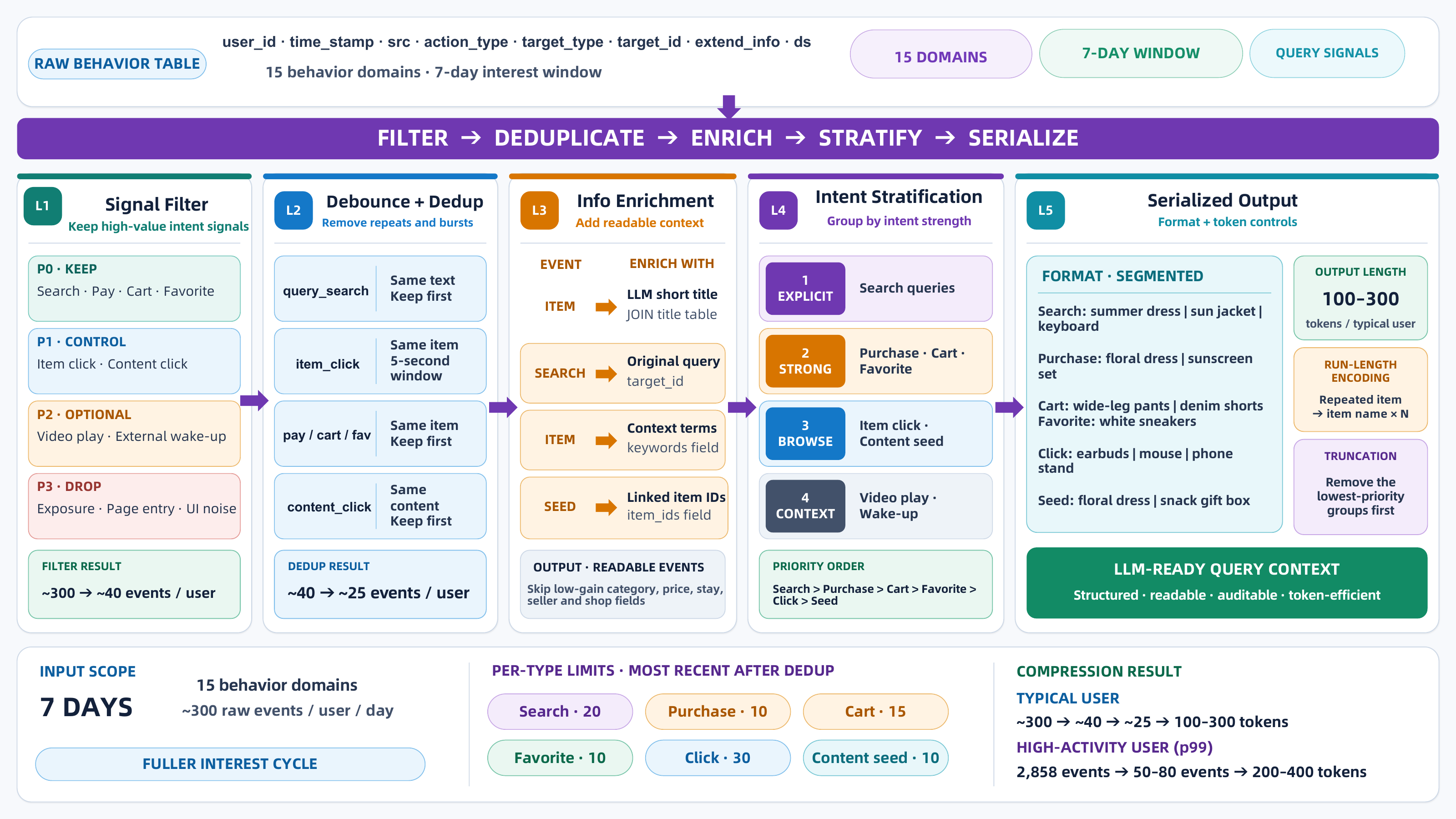}
    \caption{Five-layer behavior-trajectory compression before LLM inference: signal filtering, action-aware deduplication, semantic enrichment, intent-strength stratification, and token-efficient serialization.}
    \label{fig:behavior-compression}
\end{figure}

\begin{table}[!htbp]
    \centering
    \small
    \caption{Evidence-preserving behavior compression.}
    \label{tab:behavior-compression}
    \begin{tabularx}{\textwidth}{@{}P{0.16\textwidth}P{0.25\textwidth}YY@{}}
        \toprule
        \textbf{Layer} & \textbf{Operation} & \textbf{Retained information} & \textbf{Primary safeguard} \\
        \midrule
        Signal filter & Separate active-intent signals from passive traffic & Searches, conversions, saved items, and meaningful item/content exploration & Do not assign exposures and UI events the same evidential weight as active actions \\
        Debounce and deduplicate & Merge click bursts and repeated state changes & Unique intent-bearing actions and recurrence indicators & Preserve repeated purchase patterns while removing mechanical duplication \\
        Semantic enrichment & Attach concise product and contextual semantics & Titles, categories, selected attributes, semantic item codes, original search text & Omit fields that add little value to Query prediction or fall outside the approved field set \\
        Intent stratification & Group events by evidence strength and time scale & Explicit demand, confirmed need, browsing interest, scene context, stable preference & Truncate lower-priority context before decisive evidence \\
        Serialization & Render a compact, segmented trajectory & Short-term detail and longer-term summary in a stable schema & Version templates and audit token-level truncation \\
        \bottomrule
    \end{tabularx}
\end{table}

The serialization prioritizes semantic relevance rather than reproducing a verbatim chronological log. A Query model must identify the user's interests and strongest evidence; it rarely benefits from repeatedly processing interface transitions. The recent-activity representation preserves fine-grained intent changes, while the longer-term representation summarizes stable interests and prior relations. The two views are fused only after their distinct roles have been encoded.

The observed compression profile makes each source of information loss explicit. Signal filtering removes the majority of passive exposures and interface transitions; action-aware debouncing then collapses repeated searches, click bursts, and duplicated state changes. A typical trajectory moves from roughly three hundred raw events to about forty post-filter events and around twenty-five clean intent-bearing events. Segmented serialization usually occupies roughly one hundred to three hundred tokens, while even the high-activity tail remains within a few hundred tokens after per-event-type limits are applied. These are not training-corpus statistics; they measure the transformed context presented to the model. The decisive guardrail is that explicit searches, purchases, cart actions, favorites, and rare relation-changing items are the last to be removed.

This deterministic pipeline is the first compression boundary. Learned summarization can be added later, but it receives the cleaned representation and must reconstruct or preserve explicit searches, purchases, rare attributes, and relation-changing events. Making this boundary observable is important: the deletion of the only evidence for a valid Query is harder to diagnose when it occurs inside a latent compressor.

\subsection{Offline Learning Pipeline}

Offline learning follows a staged progression:
\begin{equation}
\begin{aligned}
    \text{General checkpoint}
    &\xrightarrow{\mathrm{general\ PT\ recovery}}
    \text{compact foundation}\\
    \text{compact foundation}
    &\xrightarrow{\mathrm{domain\ PT}}
    \text{recommendation base}\\
    \text{recommendation base}
    &\xrightarrow{\mathrm{SFT}}
    \text{Query actor}
    \xrightarrow{\mathrm{quality\mbox{-}gated\ RL}}
    \text{adaptive teacher}\\
    \text{adaptive teacher}
    &\xrightarrow{\mathrm{compress+distill}}
    \text{mobile student}.
\end{aligned}
\label{eq:offline-pipeline}
\end{equation}

Each stage has a distinct responsibility. General pre-training recovery restores broad autoregressive language-modeling capability through next-token prediction. Domain PT learns semantic identifiers, item semantics, behavioral transitions, and post-purchase relations. SFT establishes the output and reasoning contract. RL learns where reasoning can be shortened without degrading Query quality. Deployment distillation transfers both observable outputs and selected intermediate intent representations.

This progression also establishes dependencies between evaluation stages. A model that does not understand the domain cannot reliably infer intent; a model that understands the domain but lacks an output contract can produce verbose or malformed text; and a model whose quality is unstable should not yet receive an efficiency reward.

\subsection{Online Serving Pipeline}

At serving time, the device encodes the compact trajectory and uses a lightweight estimator of complexity and confidence. A request with a strong single intent is handled through direct generation. A medium-complexity request receives additional internal computation. A multi-intent or low-confidence case, or one expected to violate its budget, is escalated to a stronger service. Checks for output format, entity support, and retrieval safety apply to every final Query.

Downstream retrieval results and interaction feedback then support monitoring and data selection. Cases with large device--cloud disagreement, high predicted budgets, or meaningful differences in retrieved inventory are prioritized for review and subsequent distillation. Feedback affected by exposure or position is used only after appropriate correction; it is not treated as an immediate, unqualified reward.

\subsection{Design Principles}

Four principles keep the two layers aligned:

\begin{itemize}
    \item \textbf{Constrained efficiency.} Reasoning cost is optimized only within the set of acceptable outputs.
    \item \textbf{Instance-adaptive computation.} Budgets depend on the evidence and current policy rather than on a single global token limit.
    \item \textbf{Training--serving separation.} Explicit reasoning is retained for supervision and audit even when the deployed model uses compact internal states.
    \item \textbf{Evidence-preserving compression.} Data, reasoning, and model compression are evaluated with the same safeguards for grounding and difficult intents.
\end{itemize}

The remaining sections follow the progression in Equation~(9), beginning with the recommendation-native foundation that makes later reasoning possible.

%% file: sections/05_continual_pretraining.tex
\section{Learning a Recommendation-Native Foundation}
\label{sec:pretraining}

The model should understand behavior before it is asked to explain or optimize that behavior. We therefore use a hot-start domain adaptation stage that turns a compact general-purpose language model into a recommendation-native foundation. The objective is not to memorize a production catalog but to acquire abstractions that remain useful as inventory changes: product semantics, collaborative proximity, behavioral strength, temporal continuity, and post-purchase relations.

\subsection{Compact Hybrid Backbone and Selective Adaptation}

\begin{figure}[!htbp]
    \centering
    \includegraphics[width=\textwidth]{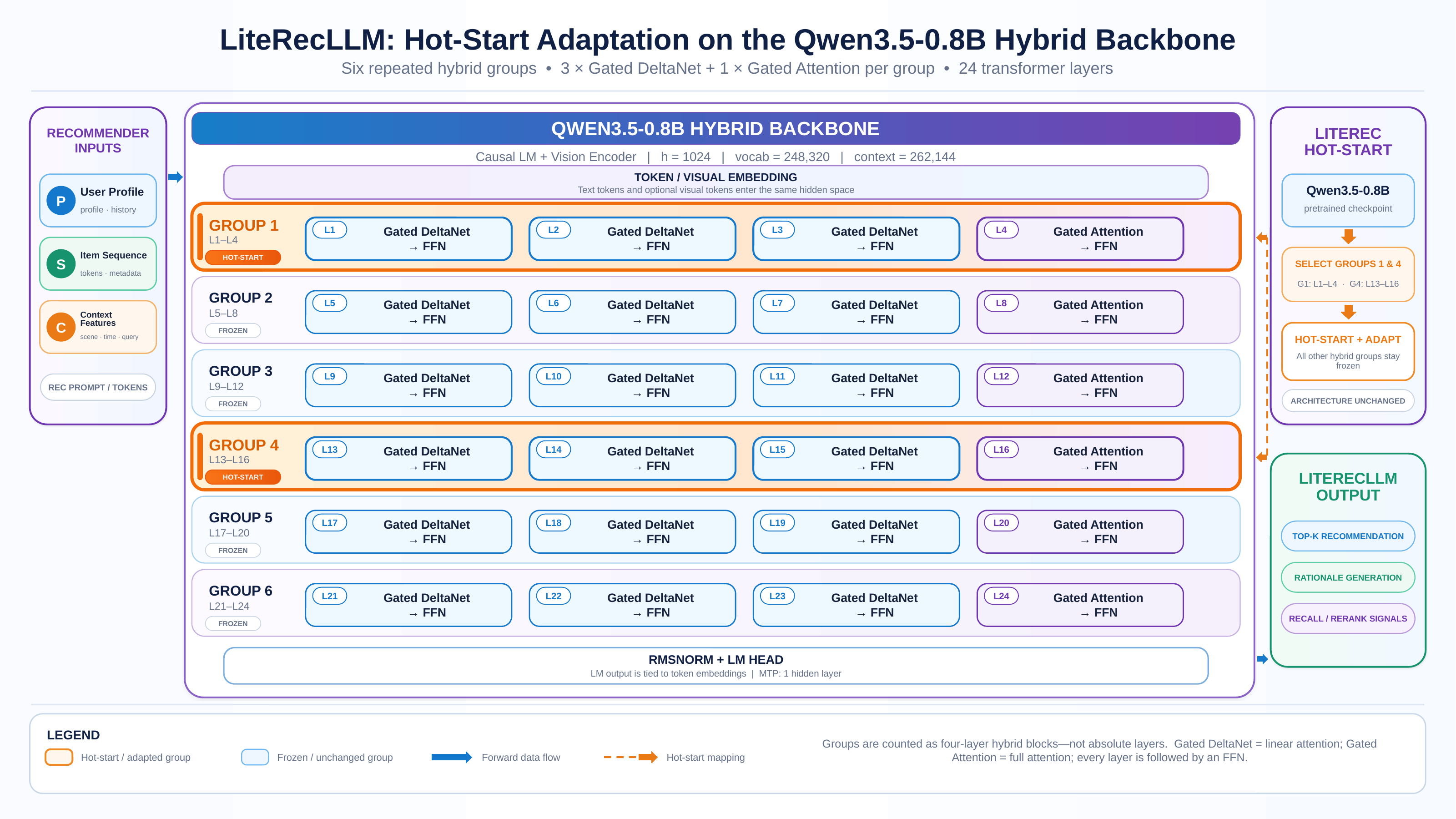}
    \caption{LiteRecLLM hot-start architecture. The compact backbone alternates efficient Gated DeltaNet mixing with periodic full attention. Selected early and middle groups are adapted to recommendation inputs, while the remaining groups preserve the general-language path.}
    \label{fig:literec-hot-start}
\end{figure}

The foundation uses the compact hybrid transformer in Figure~3. Each repeated group combines three efficient Gated DeltaNet layers with one Gated Attention layer, followed by feed-forward transformations. The linear-attention-style blocks perform most of the long-context sequence mixing; periodic full-attention layers restore global interactions among distant behaviors, profile evidence, and item semantics. This arrangement is well suited to recommendation histories because it devotes most of its computation to recurrent sequence processing while retaining regular global integration points.

Hot-start adaptation changes only selected early and middle groups. The adapted early group learns the statistics of semantic item tokens, action markers, and compact behavior serialization. The adapted middle group has a wider contextual view and specializes in modeling the interactions among short-term evidence, longer-term preference, and product relations. The other groups, the embedding interface, and the output path remain structurally intact. This selective strategy protects the general model from broad drift while giving domain-specific signals access to both local token formation and higher-level intent composition.

Selective adaptation is distinct from pruning. Hot-start adaptation specializes an architecture-preserving model. If a later deployment student removes layers, heads, or channels, it undergoes its own recovery and distillation process. Keeping these operations separate makes it possible to attribute errors to domain specialization or structural compression without conflating their effects.

\subsection{General-Capability Recovery}

Before recommendation-domain continual pre-training (CPT), we recover general capability on a governed general-domain corpus. This pre-training phase re-establishes a broad autoregressive language prior for the hot-start checkpoint without changing its topology, tokenizer, or output interface. It is neither teacher--student distillation nor task alignment and uses no Query labels, rationale targets, teacher logits, or intermediate-state supervision.

Let $w_{1:T}=(w_1,\ldots,w_T)$ denote a token sequence sampled from the general-domain corpus $\mathcal{G}$. Recovery uses the standard causal next-token prediction objective:
\begin{equation}
    \mathcal{L}_{\mathrm{recover}}(\theta)
    =-\mathbb{E}_{w\sim\mathcal{G}}
      \left[
      \frac{1}{T-1}\sum_{t=1}^{T-1}
      \log p_{\theta}\!\left(w_{t+1}\mid w_{\leq t}\right)
      \right].
    \label{eq:recovery-loss}
\end{equation}
The loss predicts every valid next token from its prefix; padding and cross-document transitions are excluded. Checkpoint selection uses held-out general-domain likelihood and fixed capability probes. The selected checkpoint initializes domain CPT, whose general stream constrains catastrophic forgetting while recommendation data introduces item semantics and behavioral relations. SFT later establishes the Query and reasoning contracts.

\subsection{Recommendation-Domain Continual Pre-Training}

The recovered checkpoint then enters recommendation-domain CPT over a governed mixture of general-domain and recommendation-domain examples. The general stream limits catastrophic forgetting, while the domain stream exposes the model to behavioral data, product language, semantic item codes, and relation-oriented tasks. We organize the domain supervision into four views:

\begin{table}[!htbp]
    \centering
    \small
    \caption{Recommendation-domain views and the capabilities they teach.}
    \label{tab:domain-views}
    \begin{tabularx}{\textwidth}{@{}P{0.24\textwidth}YY@{}}
        \toprule
        \textbf{View} & \textbf{Learning signal} & \textbf{Downstream role} \\
        \midrule
        Action semantics & Bind active behaviors to items and surfaces and encode the strength of those behaviors & Distinguish explicit demand, confirmed need, and exploratory interest \\
        Behavioral journeys & Model transitions across recent and longer-term activity & Separate transient demand from persistent preference \\
        Item and domain knowledge & Align semantic IDs with titles, categories, attributes, and use cases & Ground collaborative identity in readable product meaning \\
        Post-purchase relations & Learn relations involving repeat purchases, complementarity, substitution, co-occurrence, and scenario completion & Support Query inference that goes beyond repeating the purchased item \\
        \bottomrule
    \end{tabularx}
\end{table}

The associated training views include sequence continuation, item or semantic-code prediction, code-to-text and text-to-code alignment, relation classification, journey summarization, and teacher-guided preference targets. Abstractly,
\begin{equation}
    \mathcal{L}_{\mathrm{domain}}
    =\mathcal{L}_{\mathrm{causal}}
    +\mathcal{L}_{\mathrm{item}}
    +\mathcal{L}_{\mathrm{alignment}}
    +\mathcal{L}_{\mathrm{relation}}
    +\mathcal{L}_{\mathrm{journey}}.
    \label{eq:domain-loss}
\end{equation}
The report abstracts away corpus volumes, source-level mixture weights, and optimization settings; the scientific focus is on separating these signals and validating them on downstream tasks.

\subsection{Hierarchical Semantic Item Representation}

Opaque item identifiers provide collaborative identity but no semantic neighborhood. Titles alone are interpretable but may not capture collaborative similarity. RecGPT represents an item using a hierarchy of semantic codes together with product text and category information:
\begin{equation}
    \operatorname{item}(v)=
    [s_v^{(1)},\ldots,s_v^{(K)}]\;\Vert\;
    \operatorname{text}(v)\;\Vert\;\operatorname{category}(v).
\end{equation}
Coarse levels encode broad semantic neighborhoods; finer levels distinguish nearby items. Bidirectional tasks require the model to recover semantics from the codes and the codes from semantics. This joint representation provides four benefits:

\begin{itemize}
    \item item identity is tied to category and attribute meaning;
    \item co-occurrence, complementarity, and substitution become learnable relations;
    \item unseen or changing inventory can inherit structure from semantic neighbors; and
    \item the same token space supports behavioral modeling, retrieval, and Query generation.
\end{itemize}

Because semantic identifiers have learned meaning, their codebook and tokenizer namespace are part of checkpoint lineage. Remapping a code changes the interpretation of a trained token even if the model weights remain unchanged.

\subsection{Validation and Stage Boundary}

Pre-training validation tracks general-language retention, domain loss by view, semantic-code consistency, behavior-relation probes, and downstream Query probes. The downstream probes connect representation learning to future-intent selection and Query granularity. Checkpoint selection therefore evaluates both domain capabilities and retained performance on held-out tasks under a fixed model topology, tokenizer, and semantic-ID mapping.

Continual pre-training reshapes what the compact model can represent; it does not define the final task. The next stage supplies the missing contract: which parts of the trajectory count as evidence, what inference is allowed, and how the resulting intent must be expressed for retrieval.

%% file: sections/06_supervised_finetuning.tex
\section{Task Alignment with Structured Reasoning}
\label{sec:sft}

Domain adaptation teaches the model what behavioral evidence means; supervised fine-tuning teaches it what to do with that evidence. The central design decision is to make reasoning structured enough to support supervision and auditing, but not so elaborate that the format itself becomes the task.

\subsection{Unified Prompt and Output Contract}

All task examples use the same ordered input schema for recent behavior, longer-term history, context, and approved profile features. The target Query never appears in the prompt. SFT uses two complementary output views:

\begin{itemize}
    \item \textbf{Direct Query alignment} emits one Query without a visible rationale. It establishes the minimal serving contract and provides a no-CoT baseline.
    \item \textbf{Reasoning warm start} emits a delimited rationale followed by one final Query. It teaches behavioral attribution and intent inference before grouped policy optimization.
\end{itemize}

The final region is deliberately strict: it contains exactly one Query, with no candidate list, confidence statement, explanation, or product identifier. Reasoning and Query regions are parsed separately so that final-Query metrics are not contaminated by rationale text.

The supervised objective is ordinary masked next-token prediction:
\begin{equation}
    \mathcal{L}_{\mathrm{SFT}}(\theta)
    =-\sum_t m_t\log p_{\theta}(y_t\mid x,y_{<t}),
    \label{eq:sft-loss}
\end{equation}
where $m_t$ excludes the prompt and can distinguish rationale and final-Query regions. The distinction allows the model to learn an auditable reasoning structure without weakening the output contract.

\subsection{Five-Stage Teacher Rationale}

For supervision and analysis, a stronger teacher decomposes its rationale into five functional stages:

\begin{enumerate}
    \item \textbf{Task understanding}: identify the prediction horizon and relevant context;
    \item \textbf{Behavioral evidence}: extract the actions and entities that directly support a future need;
    \item \textbf{Candidate intents}: form a small set of plausible interpretations;
    \item \textbf{Relation judgment}: compare candidates using evidence of repeat purchases, complements, substitutions, temporal patterns, and profile attributes; and
    \item \textbf{Final decision}: select the intent and express it as a concise Query.
\end{enumerate}

This five-stage view is useful for annotation and error localization, but deployment does not require the model to verbalize all five stages. At a higher level, the rationale can be summarized as
\begin{equation}
    \text{decisive evidence}\rightarrow
    \text{candidate intent}\rightarrow
    \text{Query decision}.
    \label{eq:evidence-intent-query}
\end{equation}

\subsection{Evidence-First Short Reasoning}

The short-rationale format is not produced by truncating a long chain. It retains the stage most likely to change the answer---behavioral evidence---and a compact statement of the inferred intent. Polite restatements, exhaustive enumerations, and repeated self-confirmations are removed. Arbitrary truncation can delete the final disambiguating step; semantic compression preserves the decision path.

Training with direct and reasoned examples exposes different computational styles within a single task. Direct examples prevent visible CoT from becoming mandatory. Structured examples teach decomposition for ambiguous cases and provide intermediate targets for distillation. Together, they prepare the actor for an adaptive policy rather than a fixed response-length policy.

\subsection{Aligned CoT Information Ablation}
\label{sec:cot-ablation-contract}

We isolate the information value of reasoning with three aligned prompt conditions:

\begin{itemize}
    \item \textbf{No CoT}: the original behavioral prompt only;
    \item \textbf{Short CoT}: the original prompt plus the evidence-focused stage; and
    \item \textbf{Full CoT}: the original prompt plus the complete five-stage rationale.
\end{itemize}

Every occurrence of the observed target Query is masked from the auxiliary rationale. All three conditions share the same examples, target labels, generator, output contract, and metric implementation. This prevents label leakage and allows the comparison to answer a precise question: which intermediate information helps Query generation?

An independent large language model serves as an assessor and assigns a reasoning-need label and a diagnostic category such as direct signal, post-purchase relation, or multi-intent disambiguation. The label is used only to stratify results; it is not provided as input to the Query generator. Single-stage deletion provides a second mechanism check: if evidence extraction is genuinely central, removing that stage from the full rationale should cause the largest performance degradation.

Because the rationale is supplied as masked context in this ablation, the experiment measures information value rather than the autoregressive cost of generating that rationale. The cost of generating CoT is evaluated separately during RL and device serving. This separation is essential: a useful rationale structure is not automatically an efficient serving policy.

\subsection{Initialization and Checkpoint Selection}

The contribution of domain pre-training is tested by fine-tuning models from general and domain-adapted initializations under the same task contract. Training loss measures task-adaptation efficiency, while held-out Query similarity, behavioral grounding, output legality, relation probes, and retrieval behavior measure retained task quality and the capabilities introduced by domain learning.

SFT has completed its role when it supplies three prerequisites for adaptive reasoning: a reliable parser, enough useful candidate rollouts for group comparison, and a stable relationship between evidence and Query. Only then is it meaningful to optimize how much reasoning effort the actor expends.

%% file: sections/07_quality_gated_rl.tex
\section{Quality-Gated Adaptive Reasoning}
\label{sec:rl}

Once SFT produces reliable candidates, reinforcement learning is used to improve the quality--cost frontier. The policy follows a simple ordering: first determine whether a trajectory produces an acceptable Query; only then ask whether the same result could have been reached with less reasoning.

\begin{figure}[!htbp]
    \centering
    \includegraphics[page=1,width=\textwidth]{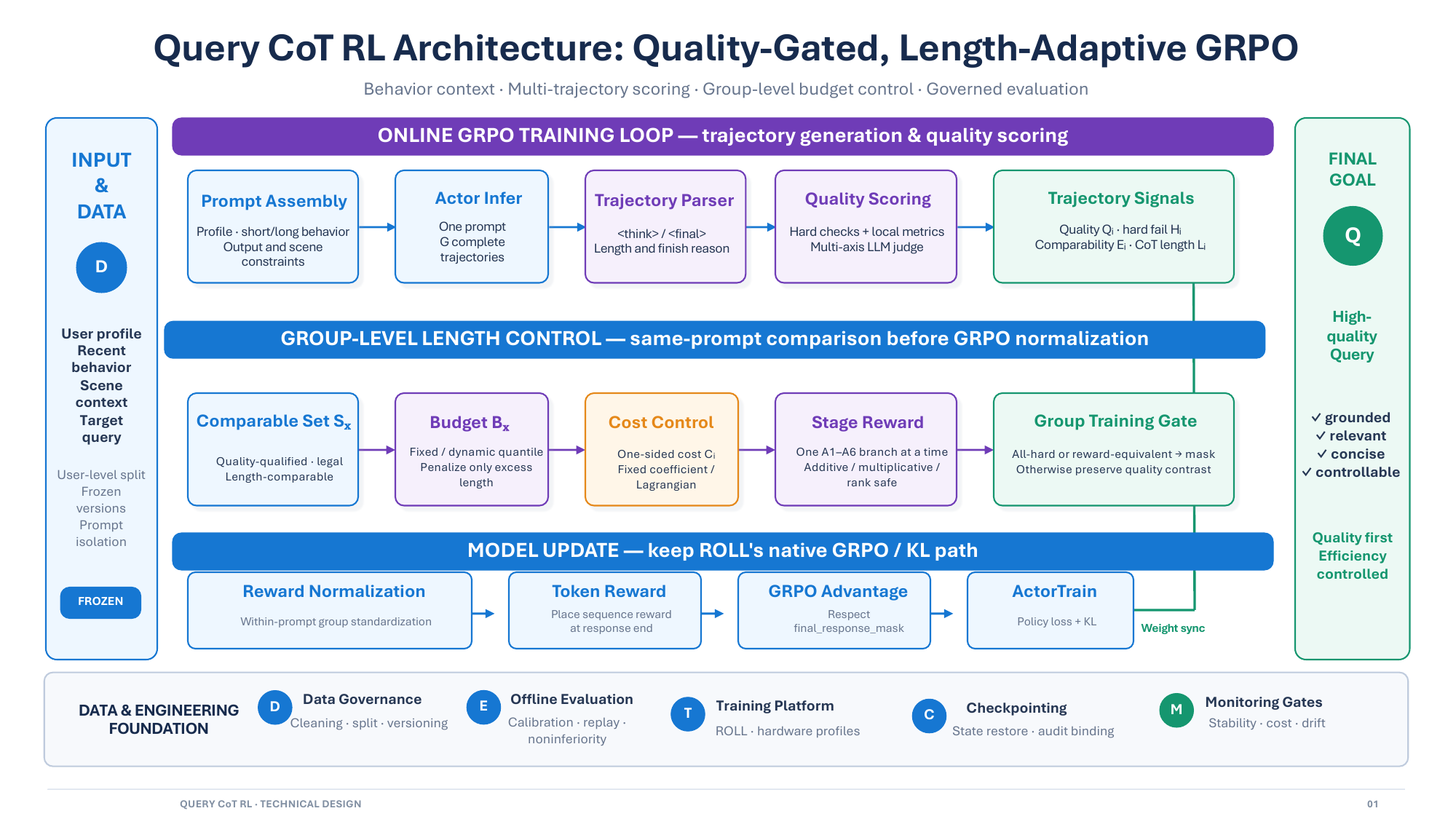}
    \caption{Query CoT RL architecture. The upper band generates and evaluates complete trajectories, the middle band compares only legal and quality-qualified trajectories from the same prompt, and the lower band preserves the native grouped policy-update path.}
    \label{fig:rl-architecture}
\end{figure}

The architecture deliberately separates three contracts. The \emph{trajectory contract} assembles the frozen behavior prompt, samples complete outputs, parses the \texttt{<think>} and \texttt{<final>} regions, and records completion status. The \emph{reward contract} combines hard checks, local text and grounding metrics, and a multidimensional model judge before any length signal is visible. The \emph{optimization contract} converts the resulting sequence reward into within-prompt advantages and applies the ordinary clipped policy objective with a reference-policy constraint. This separation lets the quality scorer, budget builder, reward shaper, and optimizer be tested independently.

Each rollout record stores the parsed Query, CoT length, finish reason, quality components, qualification and comparability masks, budget, excess cost, final reward, group advantage, and controller state. These observables are not merely implementation logs: they are the measurements needed to detect format collapse, reward-equivalent groups, quality-floor saturation, rank reversal, and unstable controller pressure.

\subsection{Why a Naive Length Penalty Fails}

A global additive length penalty confounds three situations. A short trajectory may be concise because the intent is obvious, or because the model guessed a popular Query without using the input. A long trajectory may be redundant, or it may contain a necessary cross-category inference. Since these cases have different values, length cannot be a task-independent reward.

An exact target length is equally problematic. If the model is rewarded for approaching a prescribed token count, it can fill the budget with repetition. The operational requirement is one-sided: do not exceed the amount of reasoning needed for this request. Being shorter than that amount is not independently valuable unless the Query remains good.

Fixed additive control is still informative because it can transiently reveal a low-cost direct-prediction regime with little visible rationale. The full training trajectory shows, however, that this regime is not stable without a quality gate: continued pressure eventually removes the reasoning phase faster than the policy can preserve Query grounding. We therefore evaluate the direct state by its Query quality and failure behavior, and use a separate token intervention to determine whether its residual token carries semantic content or merely preserves the output transition.

\subsection{Grouped Rollouts and Quality Qualification}

Given an input $x$, the actor samples a group of complete trajectories,
\begin{equation}
    \mathcal{Y}_x=\{(z_i,q_i)\}_{i=1}^{G}.
\end{equation}
The parser extracts the rationale, final Query, reasoning length $L_i$, completion status, and any hard-failure code. Query quality is computed without revealing $L_i$ to the evaluator. The score combines label agreement, behavioral support, and a multidimensional judgment of plausibility, specificity, naturalness, and post-purchase relevance:
\begin{equation}
    Q_i=\clip\!\left(
      w_{\ell}R_{\mathrm{label},i}
      +w_bR_{\mathrm{behavior},i}
      +w_mR_{\mathrm{model},i},0,1
    \right).
    \label{eq:rl-quality}
\end{equation}
A trajectory qualifies when its score exceeds a calibrated threshold and all hard checks pass. The threshold is tuned for precision because an invalid Query admitted to the qualified set would corrupt both the budget and the learned length preference.

\subsection{Evaluator Governance and Calibration}

The structured evaluator is a frozen Qwen3-14B model revision, served locally with deterministic decoding under a versioned evaluation contract. It receives the behavior prompt, observed label, and candidate Query, but it does not receive the generated CoT length or the reward-shaping state. The response is a strict JSON object containing seven subdimensions---behavioral anchoring, scenario coverage, pairing logic, style consistency, boundary compliance, naturalness, and length fit---plus unsupported-entity fields. These dimensions are aggregated into semantic grounding, logical validity, and language expression; semantic and logical gates prevent a fluent but unsupported Query from qualifying on expression alone. This decomposed rubric follows the general finding that structured criteria are more auditable than an undifferentiated model preference \citep{liu2023geval,zheng2023llmjudge}.

Evaluator identity is part of the experiment state. The model revision, prompt contract, Query normalizer, reward formula, and calibration artifact are frozen together; a change to any one of them creates a new scoring version. An optimizer-free calibration on development data was reviewed and frozen before evaluations of the A2--A6 mechanisms began, locking the qualification threshold, length references, budgets, and scoring versions independently of the candidate policies. Malformed JSON, an incomplete schema, a timeout after bounded retries, or mixed scoring versions fail the reward batch rather than causing a default score to be inserted. Each candidate's score is cached using a hash of the evaluator version, prompt, label, and Query, which makes repeated scoring deterministic within an experiment.

Model judgment is combined with local evidence rather than treated as ground truth. Exact match, ROUGE-L, Jaccard, deterministic output checks, and high-confidence unsupported-entity rules remain separately observable. Before a formal checkpoint is promoted, a stratified human pairwise audit measures agreement with the evaluator and the false-qualification rate across direct-signal, post-purchase, multi-intent, generic-Query, and tail-entity cases. Cases in which human preferences, lexical metrics, grounding checks, retrieval outcomes, and model-judge scores disagree form a dedicated review slice. Adversarial probes replace grounded entities with unsupported brands or specifications, substitute popular generic Queries, and add protocol-valid but irrelevant text; a candidate whose evaluator score improves while it fails these independent checks is excluded from promotion. This governance keeps the judge useful as a scalable semantic signal while limiting the reward-hacking surface exposed to the policy.

Length comparison is enabled only when the group contains at least two legal, quality-qualified trajectories with meaningful cost variation. If the group has no acceptable Query, training focuses on task quality. If it has only one such trajectory, or if all qualified trajectories have equivalent lengths, the group offers no trustworthy compression preference. Reward-equivalent groups are excluded from the policy loss because they contain no relative learning signal.

\subsection{Input-Specific Budget and One-Sided Cost}

Let $S_x$ be the qualified, length-comparable trajectories for input $x$. A robust lower quantile estimates the reasoning budget:
\begin{equation}
    B_x=\clip\!\left(
       \Quantile_{\rho}\{L_i:i\in S_x\},
       B_{\min},B_{\max}
    \right).
    \label{eq:dynamic-budget}
\end{equation}
The budget is relative to the current policy for the same input. Easy requests tend to produce several short qualified solutions and receive a smaller budget. Difficult requests either retain longer qualified solutions or fail the comparison requirement, in which case they receive no compression gradient.

Only reasoning above the budget is penalized:
\begin{equation}
    C_i=\mathbb{I}[i\in S_x]
    \frac{\max(0,L_i-B_x)}{\max(B_x,1)},
    \qquad \widehat C_i=\clip(C_i,0,C_{\max}).
    \label{eq:one-sided-cost}
\end{equation}
Further shortening within the budget earns no additional reward. This limits the incentive to delete necessary evidence or collapse toward a generic term.

\subsection{Multiplicative Reward and Rank Protection}

The efficiency term scales rather than offsets task quality:
\begin{equation}
    \widetilde R_i=Q_i\exp(-\lambda_t\widehat C_i).
    \label{eq:multiplicative-reward}
\end{equation}
The final sequence reward is
\begin{equation}
    R_i=
    \begin{cases}
      -1, & \text{hard failure},\\
      \max(\tau,\widetilde R_i), & \text{quality-qualified},\\
      Q_i, & \text{otherwise}.
    \end{cases}
    \label{eq:final-reward}
\end{equation}
Multiplicative shaping matters because an additive length bonus can compensate for a large quality deficit, while scaling can only discount quality that already exists. The floor in Equation~(20) enforces a minimum reward of $\tau$ for quality-qualified trajectories. A group-wise rank check further limits the effective length pressure whenever two trajectories have a clear quality difference. If the protected ordering cannot be maintained, the group falls back to quality-only rewards.

The coefficient $\lambda_t$ is controlled by observed excess cost rather than held fixed for the full run:
\begin{equation}
    \lambda_{t+1}=\clip\!\left(
       \lambda_t+\eta_{\lambda}(\overline C_t-\epsilon_C),
       0,\lambda_{\max}
    \right).
    \label{eq:dual-update}
\end{equation}
Here $\overline C_t$ is measured only on qualified, comparable trajectories. The controller increases pressure when the policy systematically exceeds its target and relaxes when the constraint is already satisfied. Controller state is checkpointed with the model because resetting it changes the objective even if the actor weights are identical.

\subsection{Grouped Policy Update}

Rewards are standardized within each input group,
\begin{equation}
    A_i=\frac{R_i-\mu_R(x)}{\sigma_R(x)+\epsilon},
\end{equation}
and passed to the clipped GRPO objective with a frozen reference-policy KL term \citep{shao2024deepseekmath}. The adaptive-reasoning mechanism changes the sequence reward and group mask; it does not modify token probability ratios or the underlying policy optimizer. This separation makes the reward logic independently testable and keeps the training stack comparable across ablations.

\subsection{Mechanism Ablations}

The RL study is organized as a cumulative mechanism ablation. Each arm starts from the same SFT actor and changes one design choice, as summarized in Table~3.

\begin{table}[!htbp]
    \centering
    \small
    \caption{Quality--length mechanism ablations.}
    \label{tab:rl-ablation}
    \begin{tabularx}{\textwidth}{@{}P{0.09\textwidth}P{0.34\textwidth}Y@{}}
        \toprule
        \textbf{Arm} & \textbf{Reward design} & \textbf{Question answered} \\
        \midrule
        A1 & Quality-only grouped RL & How much does post-training improve the Query before any length pressure is applied? \\
        A2 & Ungated fixed additive penalty & Can length pressure produce a high-quality direct-prediction regime, and what role does the remaining token play? \\
        A3 & Quality gate with a fixed one-sided budget & When explicit reasoning is retained, do qualification and one-sided cost protect Query quality? \\
        A4 & Input-specific dynamic budget & Does difficulty-aware budgeting improve the quality--cost frontier over one global budget? \\
        A5 & Adaptive Lagrangian coefficient & Can the policy satisfy a cost target without a manually fixed length weight? \\
        A6 & Multiplicative reward with rank protection & Does the complete reward geometry prevent short low-quality trajectories from overtaking better ones? \\
        \bottomrule
    \end{tabularx}
\end{table}

The panel is evaluated with the same prompts, actor initialization, rollout contract, Query evaluator, metric implementation, and terminal horizon. Reported metrics include mean quality, hard-failure and qualification rates, CoT length, explicit-CoT coverage, and structured semantic, logical, and expression scores. Intermediate checkpoints diagnose mechanism dynamics; the common terminal comparison supports the headline reward-design claims.

\subsection{Target Policy Behavior}

The policy is intended to learn four behaviors:

\begin{itemize}
    \item handle direct intents supported by strong evidence without unnecessary decomposition;
    \item retain evidence-comparison steps for ambiguous, multi-intent, and post-purchase relation cases;
    \item avoid repetition and unsupported specificity; and
    \item stop once the intent is stable enough to produce a grounded Query.
\end{itemize}

This creates a clean target for deployment distillation: the student learns not merely to be short, but to preserve the reasoning that changes the decision.

%% file: sections/08_compression_deployment.tex
\section{On-Device Mixed-Precision Quantization for Inference}
\label{sec:deployment}

On-device execution is constrained by parameter storage, memory bandwidth, and activation residency. We therefore focus this deployment section exclusively on mixed-precision quantization. The network topology, tokenizer, and Query policy remain unchanged. Precision is assigned at tensor granularity: insensitive operators use compact integer formats, whereas numerically fragile operators retain greater dynamic range. This isolates quantization gains from architectural changes and casts deployment as an auditable, hardware-aware allocation problem.

\subsection{Quantization Space and Tensor Granularity}

For each quantization group $g$, the design space specifies candidate precisions for weights, input activations, and key--value cache storage. Only formats supported by native kernels on the target backend are eligible. For a tensor group $X_g$ and bit width $b$, the simulated symmetric quantizer is
\begin{equation}
    \widehat{X}_{g,b}
    =s_{g,b}\operatorname{clip}
      \left(\operatorname{round}\left(X_g/s_{g,b}\right),
      -(2^{b-1}-1),\,2^{b-1}-1\right),
    \label{eq:mixed-quantizer}
\end{equation}
where $s_{g,b}$ is estimated from calibration data. Weight scales are groupwise, activation scales are computed per token or per channel according to kernel support, and cache scales are maintained per attention head. Matrix products accumulate at higher precision to prevent saturation. Embeddings, normalization operators, attention-score computation, and the output head participate in the candidate search rather than inheriting a global precision by convention.

\subsection{Sensitivity-Aware Bit Allocation}

Calibration examples are stratified by trajectory length, action diversity, tail entities, numerical attributes, direct repurchase, cross-category relations, and conflicting intents. Each candidate is first evaluated by quantizing one tensor while retaining the model at reference precision. Its sensitivity combines predictive-distribution drift with normalized hidden-state perturbation:
\begin{equation}
S_g(b)=
\lambda_p\mathbb{E}_{x\sim\mathcal{C}}
D_{\mathrm{KL}}\!\left(p_{\mathrm{fp}}(\cdot\mid x)
\Vert p_{g,b}(\cdot\mid x)\right)
+\lambda_h\mathbb{E}_{x\sim\mathcal{C}}
\frac{\|h_{\mathrm{fp}}^g-h_{g,b}^g\|_2^2}
     {\|h_{\mathrm{fp}}^g\|_2^2+\epsilon}.
\label{eq:quant-sensitivity}
\end{equation}
The groupwise choices $b$ are then obtained from
\begin{equation}
\min_{\{b_g\}}\ \sum_g C_g(b_g)
\quad\text{s.t.}\quad
\sum_g S_g(b_g)\leq \Delta,\qquad
\sum_g M_g(b_g)\leq B_m,
\label{eq:bit-allocation}
\end{equation}
where $C_g$ and $M_g$ are measured latency and memory costs. The discrete problem is solved with a knapsack-style search, followed by joint evaluation and local swaps because isolated sensitivities do not capture all cross-layer interactions. Cost tables include packing, dequantization, and precision-boundary conversions; nominal bit counts alone can otherwise favor formats that run slowly on the device.

\subsection{Calibration and Outlier Protection}

Scale selection minimizes reconstruction error after percentile clipping, with clipping thresholds chosen on the calibration split rather than from isolated extrema. Weight and activation ranges are balanced by an equivalent channel transformation,
\begin{equation}
A'_{:,j}=A_{:,j}/r_j,\qquad
W'_{j,:}=r_jW_{j,:},\qquad A'W'=AW,
\label{eq:range-balancing}
\end{equation}
where $r_j$ is selected to reduce the larger of the two quantization errors. Channels are classified as outliers only when both their calibrated magnitude and their effect on output logits exceed development thresholds. Such channels retain a higher precision; the remainder stays in the assigned low-bit format. This rule avoids preserving channels merely because they contain a single extreme observation.

\subsection{Weight, Activation, and Cache Precision}

Weights, activations, and caches require distinct treatment. Static groupwise scales are appropriate for weights, whereas changing request statistics motivate dynamic activation scales. Cache calibration spans prompt and decode positions because key and value ranges can drift as the trajectory grows. Cache outliers are protected per head, and precision is never reduced for a cache block if the resulting attention distribution violates its validation threshold.

\subsection{Quantization-Aware Recovery}

Post-training calibration supplies the initial allocation. The selected quantizers are then inserted into the forward pass, and quantization-aware recovery optimizes the same autoregressive Query objective used by the uncompressed model:
\begin{equation}
\mathcal{L}_{\mathrm{QAT}}
=-\mathbb{E}_{(x,y)}
\sum_{t=1}^{|y|}
\log p_{\theta,b}
\left(y_t\mid x,y_{<t}\right).
\label{eq:qat-recovery}
\end{equation}
Rounding is handled with a straight-through estimator, while clipping thresholds and scales remain learnable within bounded intervals. The precision assignment is frozen so that recovery cannot silently increase model cost. Training batches retain the calibration strata, and checkpoint selection uses held-out Query quality and hard-failure checks rather than weight reconstruction alone. Recovery stops when further optimization improves token loss without improving the guarded task metrics.

\subsection{Hardware and Quality Validation}

The packed artifact is profiled through the production kernels on each supported device tier. Operator coverage is checked first: an unsupported low-bit operator or an implicit floating-point fallback invalidates the measured configuration. The benchmark reports package size, peak memory, prefill and decode latency, time to first token, time per output token, throughput, energy proxy, and thermal stability under repeated requests.

Quality is compared against the reference-precision checkpoint and uniform-precision baselines under identical prompts, decoding settings, and evaluation samples. Acceptance requires preserved Query quality, output legality, behavioral grounding, and performance on tail entities, attribute-sensitive cases, long trajectories, and multi-intent inputs. The final configuration must satisfy these guardrails while improving at least one measured device cost without regressing another beyond its preset tolerance. This protocol attributes the deployment trade-off to mixed-precision quantization.

%% file: sections/09_experimental_setup.tex
\section{Experimental Design and Metrics}
\label{sec:experiments}

The evaluation is organized around causal design questions rather than a catalog of training runs. For every claim, we identify the comparison unit, the outputs that are measured, the controls that make the comparison interpretable, and the failure conditions that invalidate it. Exact optimization settings and infrastructure details remain part of the internal run record; the report focuses on model behavior and the evidence required to support each design decision.

\subsection{Experiment Matrix}

\begin{table}[!htbp]
    \centering
    \small
    \caption{Main controlled comparisons.}
    \label{tab:experiment-matrix}
    \begin{tabularx}{\textwidth}{@{}P{0.24\textwidth}P{0.31\textwidth}Y@{}}
        \toprule
        \textbf{Question} & \textbf{Comparison} & \textbf{Controlled interpretation} \\
        \midrule
        Does domain PT help task alignment? & General initialization + SFT vs. domain-adapted initialization + SFT & Separate optimization speed from held-out Query quality; verify identical model topology and task data \\
        What reasoning information matters? & No rationale vs. evidence-focused short rationale vs. full five-stage rationale & Same examples, target-leakage controls, generator, and Query metrics \\
        Which RL mechanisms matter? & A1--A6 cumulative reward ablation in Table~3 & Same SFT actor, rollouts, evaluator, and terminal horizon; one reward mechanism changes at a time \\
        Does Query prediction add recall? & Generated-Query path vs. established recall families & Compare semantic-tag overlap separately from relevance and traffic contribution \\
        Can the policy run locally? & Reference precision, uniform INT8, uniform INT4, and mixed-precision variants & Preserve Query-quality guardrails while measuring package size, memory, and latency \\
        \bottomrule
    \end{tabularx}
\end{table}

Splits are user-disjoint and the prompt template, tokenizer, semantic-ID mapping, Query normalizer, evaluator, and output parser are frozen within a controlled comparison. Intermediate checkpoints are used to diagnose training behavior; the formal reward-design panel compares arms at a common terminal horizon, and the final test is not used to tune the budget or reward.

\subsection{Behavior-Compression Evaluation}

The behavior front end is evaluated as a sequence of observable transformations rather than as one opaque prompt. For each layer, we record retained-event count, retained action families, explicit-intent recall, deduplication rate, serialized token length, and truncation by evidence tier. Manual audits focus on cases where the only search, purchase, or rare relation-bearing item would be removed. The operating profile in Figure~2 reports approximate event and token reductions for typical and high-activity users; it is interpreted together with evidence-retention checks rather than as a compression ratio alone.

\subsection{Query Quality Metrics}

No single metric fully describes a useful predicted Query. We report several complementary views:

\begin{itemize}
    \item \textbf{Lexical agreement}: exact match, BLEU, ROUGE-L, and Jaccard measure observed-label agreement at different granularities.
    \item \textbf{Semantic agreement}: a frozen encoder or judge recognizes valid paraphrases, with caps and calibration to reduce rewards for generic Queries.
    \item \textbf{Behavioral grounding}: entity and attribute extraction verifies that critical content is supported by the trajectory or label.
    \item \textbf{Task judgment}: a structured evaluator scores next-Query plausibility, post-purchase relational validity, specificity, naturalness, and output-boundary compliance.
    \item \textbf{Output distribution}: frequent-Query concentration and unique-Query ratio detect collapse toward popular expressions.
    \item \textbf{Evaluator reliability}: stratified human pairwise agreement, false-qualification rate, and adversarial score reversals audit the frozen model judge independently of policy quality.
\end{itemize}

All efficiency comparisons are conditioned on legal outputs and report quality-qualified outputs separately. This prevents a model from appearing efficient by producing empty, malformed, or irrelevant Queries.

\subsection{Domain PT and SFT Initialization Diagnostic}

The initialization study compares task-aligned models that differ in whether they begin from the general or domain-adapted foundation. Two result families are kept separate. Training loss and convergence speed describe optimization behavior; held-out Query metrics and machine-judged quality tiers describe retention. The comparison also evaluates the capabilities that domain initialization is designed to create: semantic-ID consistency, recognition of behavioral relations, and adaptation to recommendation tasks in the compact model.

\subsection{CoT Information and Stage Ablations}

The reasoning study has two complementary parts. The primary three-way comparison follows Section~6.4: No CoT, Short CoT (evidence-focused), and Full CoT (the complete five-stage rationale) are evaluated on an aligned sample set using a single metric implementation. It reports BLEU-1/2/4, ROUGE-L, and Jaccard.

A second diagnostic begins from the full five-stage rationale and removes S1--S5 one stage at a time. This diagnostic also provides the split between lower- and higher-reasoning-need inputs and the direct-signal, post-purchase-relation, and multi-intent categories. Minor record-count differences can occur after stage parsing, so this diagnostic is used to identify mechanisms and strata rather than to replace the strictly aligned three-way result. The output of the independent reasoning-need assessor is used only as a stratification variable, not as ground truth for the Query.

Because rationales are supplied as target-masked auxiliary context, this experiment is an information ablation. Latency and token cost for generated reasoning are measured only in the RL and device experiments.

\subsection{Retrieval Complementarity}

Prediction quality and retrieval novelty answer different questions. The analyzed retrieval snapshot takes the leading recalled items for each path and forms the set $T_{s,m}$ of item-level semantic tags for evaluation slice $s$ and recall path $m$. Pairwise overlap is
\begin{equation}
    J_s(m,n)=\frac{|T_{s,m}\cap T_{s,n}|}{|T_{s,m}\cup T_{s,n}|}.
    \label{eq:retrieval-jaccard}
\end{equation}
We compare the mean overlap of the Query path with established paths against the mean overlap among established paths. Lower overlap indicates complementary inventory, not relevance. Retrieval-quality checks therefore include valid-candidate count, zero-result rate, Recall@$K$, NDCG@$K$, and targeted review of the novel candidate region.

Traffic contribution and user reach are interpreted separately from overlap. A path can reach a broad set of users while still contributing a small share of exposed items; in that case, aggregate online metrics may move slowly even if the mechanism is genuinely complementary.

\subsection{Mechanism Validation and Common-Horizon Comparison}

\begin{figure}[!htbp]
    \centering
    \includegraphics[page=2,width=\textwidth]{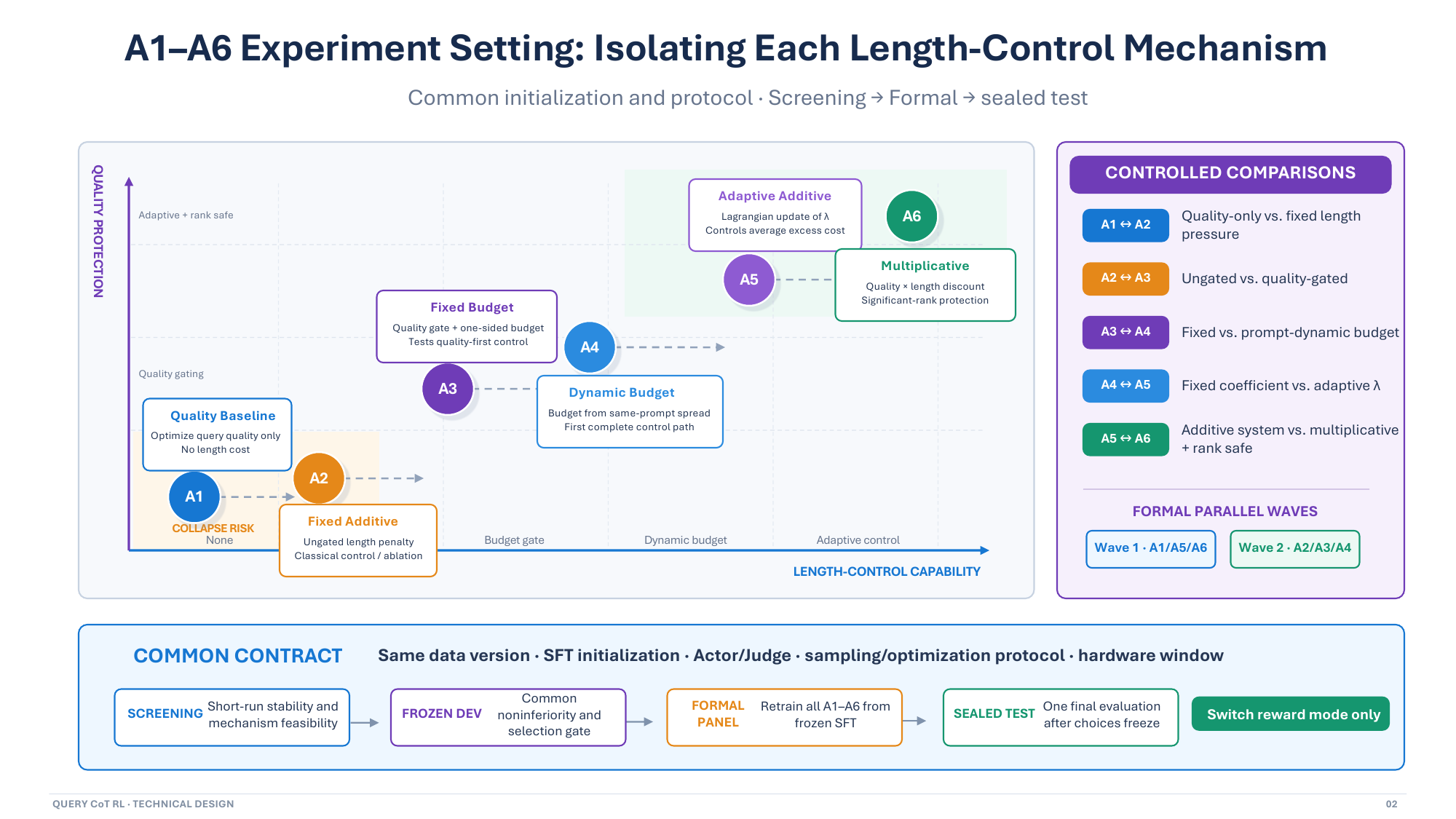}
    \caption{A1--A6 experiment map. The sequence isolates ungated length pressure, quality gating, fixed and dynamic budgets, adaptive constraint control, and multiplicative rank-safe reward shaping. Intermediate checkpoints validate the mechanism dynamics; the common terminal horizon supports the controlled comparison.}
    \label{fig:a1-a6-map}
\end{figure}

Intermediate and terminal evaluations answer different questions. Intermediate checkpoints reveal whether an arm trains stably, preserves the output protocol, and exposes a well-characterized operating regime. A dedicated intervention on the one-token state separates the effect of token presence from token identity. The common terminal horizon then shows whether that behavior remains stable under continued optimization.

In the formal A1--A6 panel, every arm is restarted from the same SFT actor, and the prompt set, evaluator, rollout policy, and metric implementation are frozen. Intermediate checkpoints expose training dynamics, while the headline comparison evaluates every arm at the same terminal horizon. This prevents an unstable early shortcut from being compared with a fully optimized constrained policy and makes each cumulative mechanism change interpretable.

\subsection{RL Quality--Cost Metrics}

The A1--A6 panel reports:

\begin{itemize}
    \item mean Query quality, qualification rate, hard-failure rate, and format-failure rate;
    \item median and tail CoT lengths together with explicit-CoT coverage;
    \item checkpoint-level quality and length dynamics; and
    \item decomposed semantic, logical, expression, and weighted quality scores.
\end{itemize}

The primary output is a quality--cost Pareto frontier rather than a single token number. A method is useful if it reduces mean and tail reasoning lengths while preserving aggregate quality and performance on difficult intent strata. Comparisons at approximately matched mean CoT lengths separate reward geometry from compression strength.

\subsection{Device Metrics}

The serving benchmark reports package size, peak memory, time to first token, time per output token, end-to-end latency distribution, throughput, energy proxy, thermal behavior, kernel coverage, and precision-boundary conversion overhead. Reference precision, uniform INT8, uniform INT4, and mixed-precision assignments are evaluated under identical runtime and device controls, allowing measured improvements to be attributed to precision allocation.

\subsection{Statistical and Acceptance Protocol}

Models are evaluated pairwise on the same samples. Confidence intervals use user-clustered resampling, and key difficult strata are preregistered before the final comparison. Human review prioritizes disagreements among textual metrics, the structured evaluator, and retrieval behavior; large Query changes; multi-intent cases; conflicting evidence; and tail entities.

A candidate is accepted only if it preserves Query-quality guardrails, does not increase hard failures or unsupported entities, avoids popular-Query collapse, reduces the relevant reasoning cost or device cost, and remains stable across difficult strata. Lower retrieval overlap is considered beneficial only when relevance checks also pass.

%% file: sections/10_results_analysis.tex
\section{Results and Analysis}
\label{sec:results}

The experiments are organized as a chain of evidence. Behavior compression first establishes that useful intent can be exposed within a bounded context. The initialization study asks whether domain adaptation changes downstream learning. The CoT experiments identify which intermediate information improves the Query and for which inputs. Retrieval analysis tests whether those Queries reach a different item region. Finally, the controlled RL panel tests whether the model can retain that quality while reducing generated reasoning, and the serving benchmark measures how the shorter policy interacts with deployment optimization. Keeping these questions separate prevents a favorable number at one stage from being mistaken for end-to-end value.

\subsection{Behavior Compression Produces a Stable Context Envelope}

Table~5 summarizes the operating profile shown in Figure~2. Filtering removes passive exposures, page transitions, and low-value interface events; action-specific deduplication then removes repeated searches, burst clicks, and duplicate state changes. The large reduction occurs before any learned summarizer, so the discarded information remains auditable.

\begin{table}[!htbp]
    \centering
    \small
    \caption{Observed behavior-compression profile. Values are approximate operating measurements.}
    \label{tab:behavior-compression-profile}
    \begin{tabularx}{0.92\textwidth}{@{}P{0.24\textwidth}P{0.20\textwidth}Y@{}}
        \toprule
        \textbf{Representation point} & \textbf{Typical profile} & \textbf{What remains} \\
        \midrule
        Raw behavior stream & $\sim$300 events & Cross-surface exposures, navigation, active actions, and repeated interactions \\
        After signal filtering & $\sim$40 events & Search, purchase, cart, favorite, controlled browsing, and selected context \\
        After debouncing/deduplication & $\sim$25 events & Unique intent-bearing actions with readable item and Query semantics \\
        Segmented LLM context & $\sim 100$--$300$ tokens & Evidence grouped by intent strength rather than repetitive timestamped prose \\
        High-activity tail & $50$--$80$ events; $\sim 200$--$400$ tokens & Evidence capped by type, with the lowest-priority groups truncated first \\
        \bottomrule
    \end{tabularx}
\end{table}

The resulting context is short enough to make model-side reasoning the dominant variable, but it is not simply the shortest possible prompt. Search terms and conversion actions are retained ahead of browsing and contextual events, and the serialization keeps short- and longer-term evidence distinguishable. This matters for the later ablations: if decisive behavior had already been removed, a CoT experiment would measure prompt damage rather than reasoning value.

\subsection{Domain Initialization Establishes an Optimization--Retention Trade-off}
\label{sec:sft-results}

The model with domain-adapted initialization reaches a lower task-training loss than the model with general initialization in the SFT comparison. Mean loss decreases from $0.6448$ to $0.6066$, a relative reduction of approximately $5.9\%$, with the largest separation early in training. This result shows that the Query contract is easier to learn from a model that already represents semantic IDs, behavior transitions, and recommendation-domain relations.

The purpose of this initialization is broader than increasing lexical overlap with one observed label. It establishes the semantic-ID interface, priors over behavioral relations, and the compact recommendation backbone used by the subsequent stages. The held-out evaluation shows that CPT preserves the overall machine-judged Query score at $99.58\%$ of Base+SFT while increasing the highest-quality A-tier share from $39.22\%$ to $39.70\%$. CPT retains approximately $95$--$96\%$ of Base+SFT performance on the lexical metrics. Together with the lower task-training loss, these results show that CPT provides a useful recommendation-domain starting point without materially changing the overall quality profile.

\begin{figure}[!htbp]
    \centering
    \includegraphics[width=0.94\textwidth]{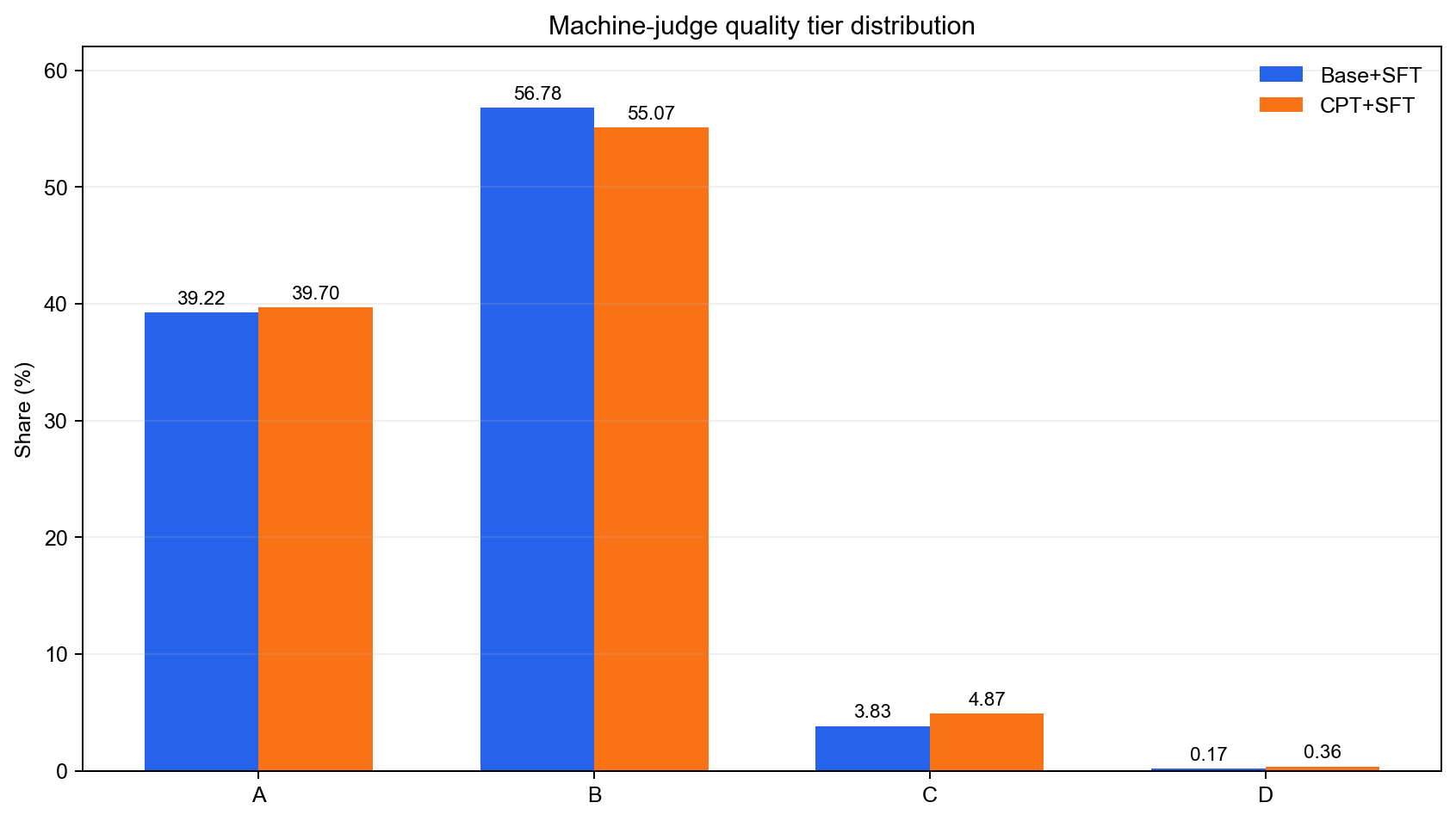}
    \caption{Machine-judged quality-tier distribution after SFT. CPT preserves a quality profile close to Base+SFT while slightly increasing the A-tier share.}
    \label{fig:sft-analysis}
\end{figure}

Taken together, the lower training loss and stable held-out quality establish a practical optimization--retention trade-off. Domain PT gives the smaller, deployment-oriented model a recommendation-native interface without requiring SFT to relearn item semantics and behavioral relations from scratch. Its benefit is therefore measured through easier task adaptation, preserved Query quality, semantic-ID consistency, relation understanding, and serving efficiency rather than through lexical overlap alone.

\subsection{What CoT Adds: Evidence and Selective Benefit}
\label{sec:cot-results}

The primary reasoning experiment compares three aligned views of the same task examples. No CoT provides only the behavior prompt. Short CoT adds a target-masked behavioral-evidence stage. Full CoT adds the complete five-stage rationale. All three share the generator, final-Query contract, sample alignment, and metric implementation.

\begin{table}[!htbp]
    \centering
    \small
    \caption{Aligned information ablation for No CoT, Short CoT, and Full CoT.}
    \label{tab:cot-three-way}
    \begin{tabular}{lrrrrr}
        \toprule
        \textbf{Prompt view} & \textbf{BLEU-1} & \textbf{BLEU-2} & \textbf{BLEU-4} & \textbf{ROUGE-L} & \textbf{Jaccard} \\
        \midrule
        No CoT & 0.2094 & 0.1486 & 0.0457 & 0.2281 & 0.1740 \\
        Short CoT: evidence-focused & \textbf{0.2925} & \textbf{0.2200} & \textbf{0.0767} & \textbf{0.3147} & \textbf{0.2481} \\
        Full CoT: complete S1--S5 & 0.2909 & 0.2186 & 0.0756 & 0.3098 & 0.2441 \\
        \bottomrule
    \end{tabular}
\end{table}

Short CoT raises ROUGE-L by $0.0866$ and Jaccard by $0.0741$ over direct prediction, while slightly exceeding the full rationale. The important result is not that long reasoning is intrinsically harmful; it is that behavioral evidence extraction accounts for most of the measurable benefit, whereas additional narrative stages provide little aggregate gain.

\begin{figure}[!htbp]
    \centering
    \includegraphics[width=\textwidth]{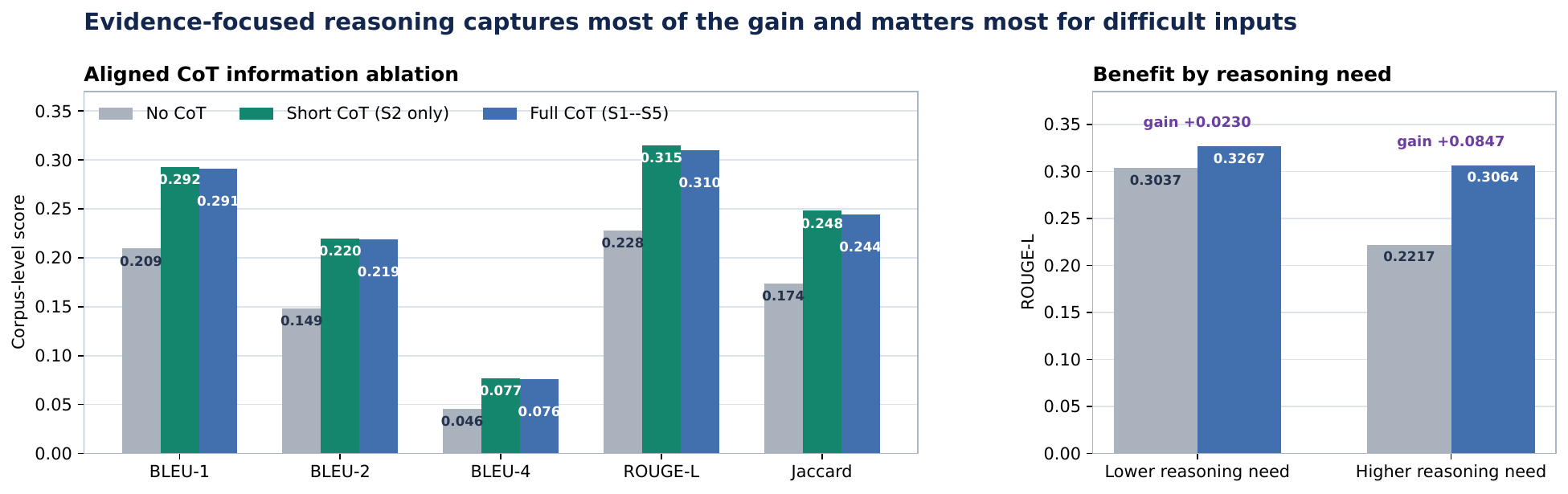}
    \caption{The two decision-relevant CoT results. Left: evidence-focused Short CoT retains the gain of the complete rationale. Right: full-rationale information provides a much greater benefit for inputs independently assessed as requiring reasoning.}
    \label{fig:cot-three-way}
\end{figure}

Two diagnostics support this interpretation without requiring a separate figure for every deletion. First, removing evidence extraction from the full rationale lowers ROUGE-L from $0.3077$ to $0.2806$ and Jaccard from $0.2415$ to $0.2177$, representing the largest consistent stage-level deterioration. Removing candidate enumeration or explicit relation narration instead changes aggregate scores only slightly and can even improve them, indicating redundant verbalization rather than missing evidence.

Second, an independent large language model assessor stratifies inputs without exposing the resulting label to the Query generator. Full CoT improves ROUGE-L by only $0.0230$ on lower-need inputs but by $0.0847$ on higher-need inputs. The same concentration of gains appears in the task categories below.

\begin{table}[!htbp]
    \centering
    \small
    \caption{ROUGE-L gain from full CoT by input category.}
    \label{tab:cot-reason-gains}
    \begin{tabular}{lrrr}
        \toprule
        \textbf{Input category} & \textbf{No CoT} & \textbf{Full CoT} & \textbf{Gain} \\
        \midrule
        Direct signal & 0.3024 & 0.3256 & $+0.0232$ \\
        Post-purchase relation & 0.2241 & 0.3089 & $+0.0848$ \\
        Multi-intent disambiguation & 0.1047 & 0.1914 & $+0.0867$ \\
        \bottomrule
    \end{tabular}
\end{table}

The practical conclusion is straightforward: preserve the evidence stage, make further reasoning conditional on ambiguity, and compress narrative elaboration first. Because the rationales in this experiment are supplied as target-masked auxiliary context, the comparison measures their \emph{information value}; generated reasoning cost is evaluated separately in the RL study.

\subsection{The Query Path Retrieves Complementary Inventory}

The retrieval analysis compares semantic-tag sets among established recall paths and the generated-Query path. Figure~8 shows four evaluation slices. The solid red line plots the Query row from each pairwise Jaccard matrix. It separates from the dense cluster formed by established paths in three slices and shows comparable overlap in the fourth.

\begin{figure}[!htbp]
    \centering
    \includegraphics[width=\textwidth]{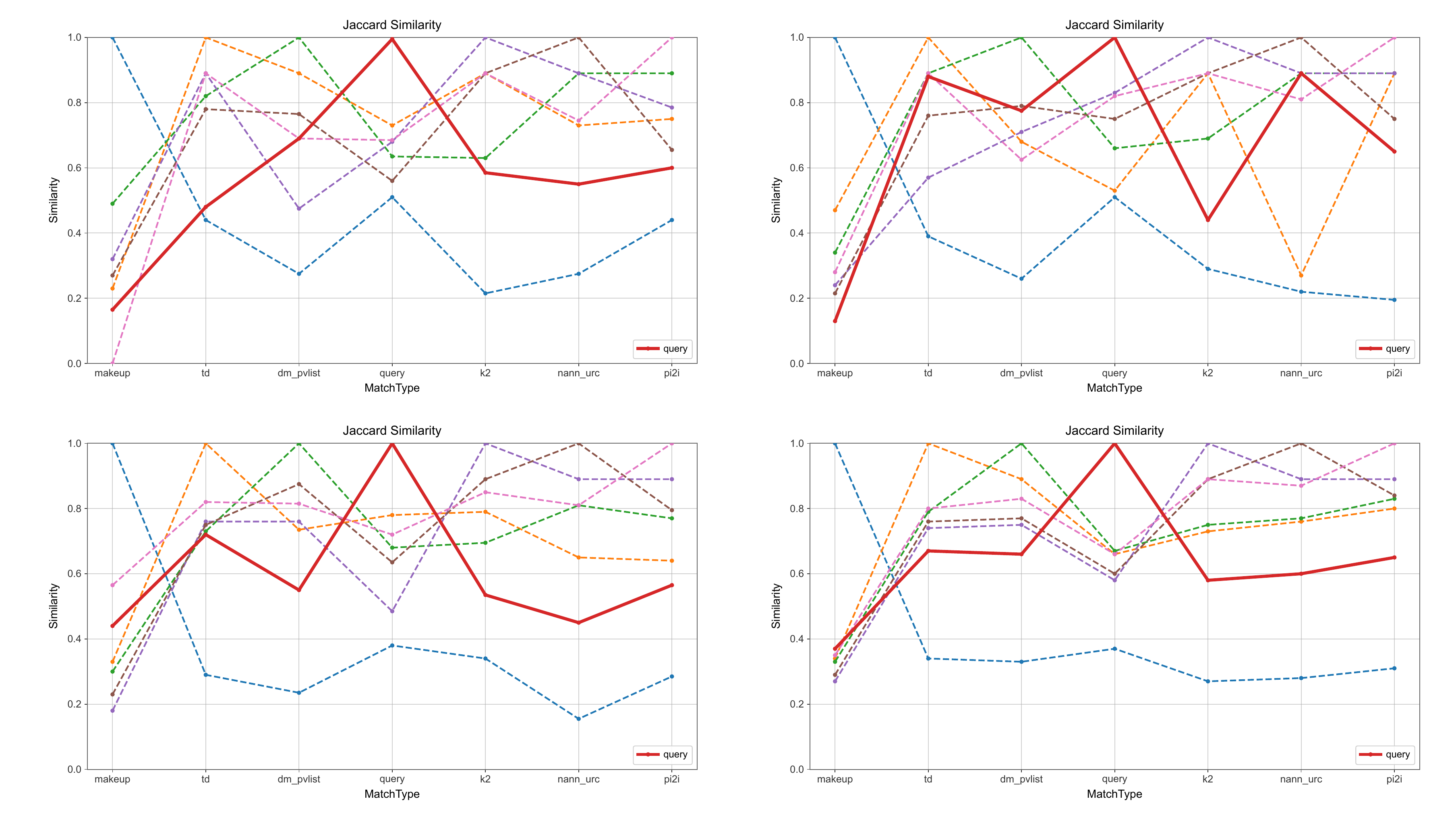}
    \caption{Pairwise Jaccard similarity of semantic item tags among leading recalled candidates. The emphasized red curve corresponds to the generated-Query path. Lower cross-path overlap indicates a different semantic inventory, not automatically higher relevance.}
    \label{fig:tagv5-jaccard}
\end{figure}

Across the four slices, the mean Query-to-other overlap values are $0.512$, $0.628$, $0.543$, and $0.588$, compared with $0.626$, $0.615$, $0.621$, and $0.645$ for pairs of established paths. The Query path therefore shows lower overlap in three slices, by approximately $0.06$--$0.12$, and is approximately tied in the fourth. The repeated separation indicates that behavior-derived Queries reach a semantic region not fully covered by the dominant recall families.

The contribution profile in Table~8 provides context for the scale of this effect. Server-side Query recall accounts for a small fraction of upstream recall and exposed items while reaching a materially larger fraction of users. This combination is consistent with a broad, complementary source whose incremental inventory is distributed across the user population.

\begin{table}[!htbp]
    \centering
    \small
    \caption{Recall-path contribution by serving scope.}
    \label{tab:path-contribution}
    \begin{tabular}{llrrr}
        \toprule
        \textbf{Scope} & \textbf{Path} & \textbf{Recall share} & \textbf{Exposed-PV share} & \textbf{UV coverage} \\
        \midrule
        Server & k2 & 8.00\% & 67.00\% & 38.90\% \\
        Server & nann\_urc & 12.00\% & 19.00\% & 25.40\% \\
        Server & pi2i & 5.00\% & 4.70\% & 11.30\% \\
        Server & makeup & 67.00\% & 1.70\% & 1.10\% \\
        Server & dm\_pvlist & 1.00\% & 1.70\% & 6.50\% \\
        Server & td & 0.60\% & 1.30\% & 5.10\% \\
        Server & \textbf{Query} & \textbf{0.20\%} & \textbf{1.10\%} & \textbf{6.90\%} \\
        \midrule
        Client & Query Android & 0.05\% & 0.12\% & 0.70\% \\
        Client & Query iOS & 0.06\% & 0.18\% & 1.20\% \\
        \bottomrule
    \end{tabular}
\end{table}

The operational sequence is clear. First, improve Query grounding, candidate validity, and relevance within the existing opportunity. Next, verify that the low-overlap region contributes useful rather than merely different items. Only then should allocation be increased. Jaccard quantifies complementarity; it does not replace Recall@$K$, NDCG@$K$, user feedback, or controlled traffic evaluation.

\subsection{RL Training Dynamics Validate the Reward Mechanisms}
\label{sec:rl-formal-results}

The controlled panel begins each arm from the same policy state and evaluates every reward design over a common training horizon. Figure~9 separates Query quality from reasoning cost so that a short but degraded endpoint cannot be mistaken for an efficiency gain.

\begin{figure}[!htbp]
    \centering
    \includegraphics[width=\textwidth]{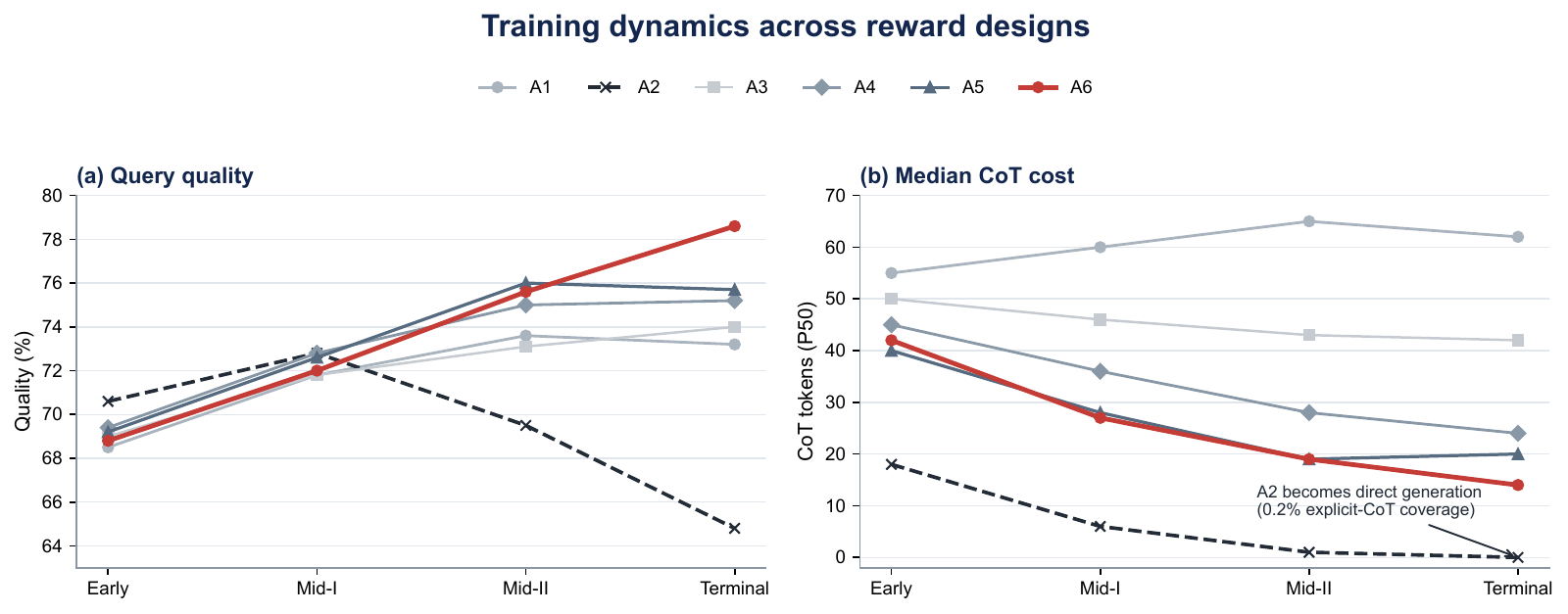}
    \caption{Training dynamics for A1--A6. The left panel tracks Query quality; the right panel tracks median CoT length. A2 is shown as a direct-generation trajectory because its explicit-CoT coverage collapses as training proceeds.}
    \label{fig:rl-training-dynamics}
\end{figure}

A1 establishes that quality-only GRPO is effective but not efficient by itself. Relative to A0, terminal Query quality rises from $65.5\%$ to $73.2\%$, the qualification rate rises from $48.0\%$ to $64.5\%$, and the hard-failure rate falls from $8.5\%$ to $3.6\%$. Median CoT length nevertheless increases from $52$ to $62$ tokens. Quality peaks at $73.6\%$ before a small terminal regression, while the reasoning path remains longer than the baseline. The policy has learned a better answer strategy, but it has not learned when additional reasoning stops paying off.

A2 exposes the shortcut created by an ungated additive length penalty. Its early trajectory appears favorable: quality reaches $72.8\%$ while median CoT falls to $6$ tokens. Continued optimization, however, drives explicit-CoT coverage from $72.0\%$ to $0.2\%$ and terminal quality to $64.8\%$; the hard-failure rate rises to $10.5\%$. The zero-token endpoint is therefore a direct-generation regime whose cost cannot be interpreted independently of its Query degradation. This is a failure in reward semantics rather than evidence that direct generation is intrinsically poor.

A separate intervention clarifies the transient one-token state observed before the A2 endpoint. Under the same two-stage output protocol, deleting the sampled transition token reduces Query quality by $11.08$ percentage points, increases the hard-failure rate by $15.53$ points, and reduces the qualification rate by $9.47$ points. Replacing it with a shuffled token changes the same metrics by only $+0.23$, $-0.05$, and $+0.34$ points. The token therefore functions mainly as a boundary between internal processing and final generation, not as a one-token semantic summary. This distinction allows the direct regime to be analyzed in terms of Query quality rather than being credited merely for brevity.

A3 repairs the A2 incentive by applying length pressure only to quality-qualified trajectories and only above a fixed budget. Terminal quality recovers by $9.2$ points to $74.0\%$, the qualification rate recovers by $25.4$ points to $67.5\%$, and the hard-failure rate falls by $7.5$ points to $3.0\%$. Explicit-CoT coverage returns to $97.4\%$, confirming that the recovery comes from restoring quality-conditioned reasoning rather than from a format workaround.

A4--A6 then improve the same qualified regime rather than changing the output contract. The input-specific budget produces the largest isolated cost reduction: A4 cuts CoT P50 from $42$ to $24$ tokens while raising quality by $1.2$ points. The adaptive controller in A5 reduces P50 further to $20$ and raises quality to $75.7\%$, although it shows a small late-stage reversal in both quality and length near the constraint boundary. A6 starts deliberately slowly but continues improving through the terminal horizon: quality increases from $68.8\%$ to $78.6\%$ while P50 falls from $42$ to $14$ tokens. This delayed payoff is consistent with rank protection first stabilizing the trajectory ordering and then compressing the winning behavior.

\subsection{Controlled Common-Horizon Comparison}

\input{tables/formal_rl_results}

\begin{figure}[!htbp]
    \centering
    \includegraphics[width=\textwidth]{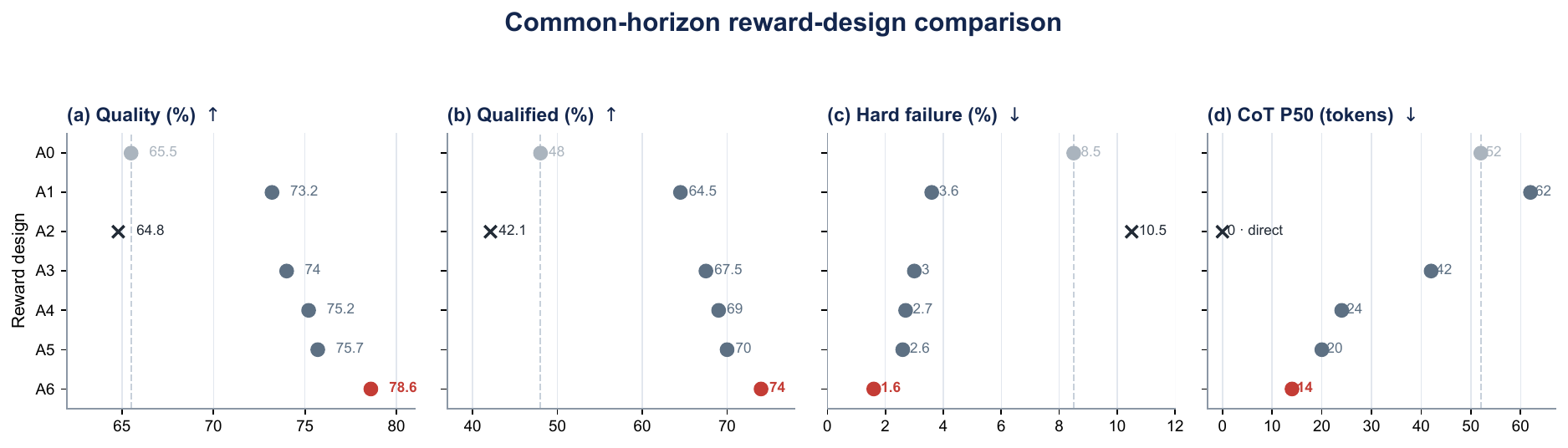}
    \caption{Terminal mechanism comparison. Dashed lines mark the A0 baseline. A2 is a direct endpoint with almost no explicit CoT; A6 provides the best quality--cost trade-off among the quality-preserving reasoning designs.}
    \label{fig:rl-terminal-comparison}
\end{figure}

The terminal comparison makes the cumulative contribution of the mechanisms explicit. A3 proves that quality gating is the essential safeguard. A4 shows that replacing one global budget with a per-input budget is not a cosmetic change: it reduces the median CoT length by $18$ tokens while improving every task-quality indicator. A5 shows that constraint strength can be learned, but its small late rebound reveals the stability limit of a purely additive controller. A6 removes the remaining compensation path by making length gains proportional to existing quality and preserving clear quality orderings.

A6 reaches $78.6\%$ mean Query quality, a $74.0\%$ qualification rate, a $1.6\%$ hard-failure rate, and a $14$-token CoT median. Relative to the quality-only A1 policy, it gains $5.4$ quality points and $9.5$ qualification points, reduces the hard-failure rate by $2.0$ points, and reduces the median reasoning length by $48$ tokens---a $77.4\%$ reduction. Relative to A0, it gains $13.1$ quality points while reducing the CoT median by $73.1\%$. Its $98.5\%$ explicit-CoT coverage is important: unlike A2, the shorter endpoint is not produced by abandoning the reasoning protocol.

\subsection{Structured Quality Explains the A6 Gain}

The aggregate result is supported by the decomposed rubric scores in Table~10 and Figure~11. The evaluator separates semantic grounding and post-purchase logic from expression quality, which is necessary for detecting fluent reward shortcuts.

\begin{figure}[!htbp]
    \centering
    \includegraphics[width=\textwidth]{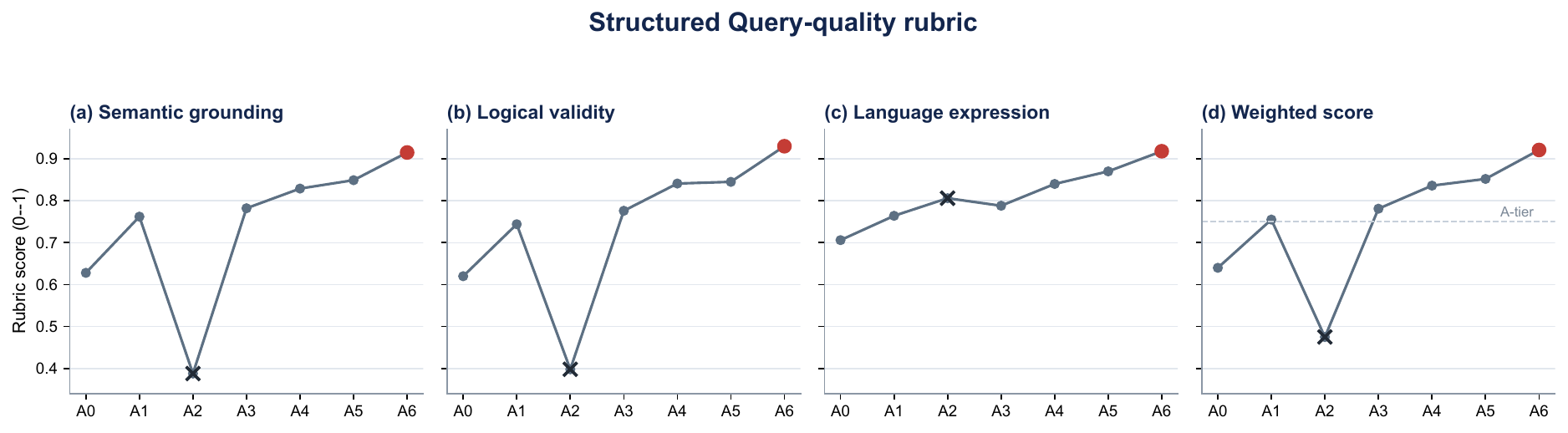}
    \caption{Structured Query-quality diagnosis. A2 retains fluent search-like expression while losing semantic grounding and post-purchase logic; A3 restores both, and A4--A6 progressively improve performance across the complete rubric.}
    \label{fig:rl-rubric-ablation}
\end{figure}

A2 is the decisive diagnostic case. Its expression score is $0.806$, higher than A1's $0.764$, even though semantic grounding and logical validity fall to $0.388$ and $0.398$. The policy has learned to emit concise strings that look like Queries, but those strings are no longer reliably supported by the behavior trajectory. A quality signal that collapsed these dimensions into surface naturalness would incorrectly reward this endpoint.

Quality gating in A3 restores the semantic and logical scores to $0.782$ and $0.776$. Dynamic budgeting in A4 then raises them to $0.829$ and $0.841$ while shortening the rationale, showing that request-specific allocation improves reasoning selection rather than merely reducing text. A6 reaches $0.915$ in semantic grounding, $0.930$ in logical validity, and $0.918$ in expression quality, with a weighted score of $0.921$. The gain is therefore distributed across evidence, relation judgment, boundary control, and final expression rather than concentrated in one lexical metric.

\subsection{Serving Consequence of the Shorter Policy}

The effect of reasoning compression is amplified after model and runtime optimization. Figure~12 reports the end-to-end comparison; it should be interpreted as the combined result of a shorter output policy and a compact serving stack, not as an RL-only speedup.

\begin{figure}[!htbp]
    \centering
    \includegraphics[width=0.75\textwidth]{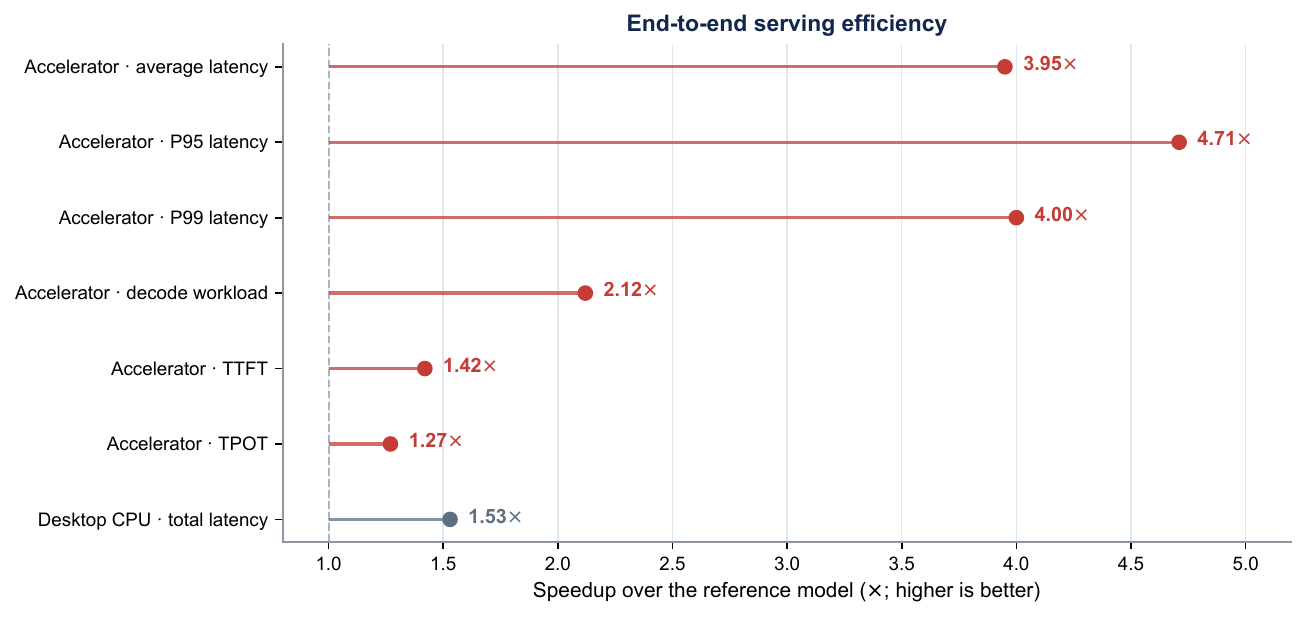}
    \caption{End-to-end serving speedup over the reference model. Shorter decoding contributes directly, while model compression and runtime optimization further improve per-token time and tail latency.}
    \label{fig:serving-efficiency}
\end{figure}

On the accelerator path, average latency decreases from $3.00$ to $0.76$ seconds ($3.95\times$), P95 from $8.00$ to $1.70$ seconds ($4.71\times$), and P99 from $10.00$ to $2.50$ seconds ($4.00\times$). Performance on the isolated decode workload improves by $2.12\times$, while time to first token and time per output token improve by $1.42\times$ and $1.27\times$. On the desktop CPU path, total latency decreases from $4.13$ to $2.70$ seconds ($1.53\times$). The larger tail-latency gain indicates that removing unnecessarily long generations reduces scheduling variability in addition to lowering the average decoding workload.

\subsection{Combined Interpretation}

The experiments support six connected conclusions.

\begin{enumerate}
    \item \textbf{Compression should expose evidence, not merely shorten text.} The deterministic front end removes most passive traffic while preserving explicit-intent and conversion actions in a bounded context.
    \item \textbf{Domain adaptation is an optimization--retention trade-off.} It makes task alignment easier and adds a recommendation-native representation while retaining most held-out Query quality; downstream selection jointly considers task quality, semantic-ID and relation probes, and serving cost.
    \item \textbf{Behavioral evidence is the core of useful CoT.} Short evidence-focused CoT matches or exceeds a full narrative, and deleting evidence extraction causes the largest stage-level loss.
    \item \textbf{Reasoning should be conditional.} Direct signals gain little from long CoT, whereas post-purchase relations and multi-intent disambiguation gain substantially.
    \item \textbf{Length optimization must be conditioned on quality.} Quality-only RL improves the Query but lengthens reasoning; ungated additive pressure creates a fluent direct-generation shortcut; per-input budgets provide the largest isolated cost reduction; and multiplicative rank-safe shaping produces the strongest joint endpoint.
    \item \textbf{Efficiency claims need causal and downstream checks.} A one-token state can be a useful output transition without being a compressed textual rationale; low retrieval overlap can indicate novelty without guaranteeing relevance. Interventions, decomposed quality scores, analysis of difficult strata, and retrieval-relevance checks must accompany every efficiency or diversity claim.
\end{enumerate}

The empirical evidence therefore supports neither the claim that \emph{long reasoning is always better} nor the claim that \emph{short reasoning is always better}. The appropriate design uses a compact evidence representation, followed by additional computation only when relational or ambiguous evidence can change the final Query. In this setting, A6 is not merely the shortest qualified policy: it improves the Query while shortening the reasoning path, which is the defining property of a better quality--cost frontier.

%% file: tables/formal_rl_results.tex
\begin{table}[!htbp]
  \centering
  \caption{Common-horizon terminal comparison of the A0--A6 reward designs. Quality, qualification, and failure values are percentages.}
  \label{tab:formal-rl-quality}
  \scriptsize
  \setlength{\tabcolsep}{3.8pt}
  \begin{tabularx}{\textwidth}{@{}lYrrrrrr@{}}
    \toprule
    \textbf{Arm} & \textbf{Reward design} & \textbf{Quality} & \textbf{Qualification} & \textbf{Hard failure} & \textbf{Format failure} & \textbf{Explicit CoT} & \textbf{CoT P50} \\
    \midrule
    A0 & Current policy baseline & 65.5 & 48.0 & 8.5 & 2.50 & 92.0 & 52 \\
    A1 & Quality-only grouped RL & 73.2 & 64.5 & 3.6 & 0.12 & 97.0 & 62 \\
    A2 & Ungated additive length penalty & 64.8 & 42.1 & 10.5 & 0.90 & 0.2 & 0 \\
    A3 & Quality gate + fixed one-sided budget & 74.0 & 67.5 & 3.0 & 0.07 & 97.4 & 42 \\
    A4 & Input-specific dynamic budget & 75.2 & 69.0 & 2.7 & 0.20 & 97.0 & 24 \\
    A5 & Adaptive Lagrangian coefficient & 75.7 & 70.0 & 2.6 & 0.15 & 97.6 & 20 \\
    \textbf{A6} & \textbf{Multiplicative reward + rank protection} & \textbf{78.6} & \textbf{74.0} & \textbf{1.6} & \textbf{0.04} & \textbf{98.5} & \textbf{14} \\
    \bottomrule
  \end{tabularx}
  \vspace{0.3em}
  \begin{minipage}{0.97\textwidth}
    \footnotesize A2's zero-token median denotes a direct-generation endpoint, not a quality-preserving CoT compression point; its explicit-CoT coverage, Query quality, and failure rate must be read jointly.
  \end{minipage}
\end{table}

\begin{table}[!htbp]
  \centering
  \caption{Structured Query-quality diagnosis at the common terminal horizon. The weighted score aggregates semantic grounding, logical validity, and expression quality.}
  \label{tab:formal-rl-rubric}
  \small
  \setlength{\tabcolsep}{8.0pt}
  \begin{tabular}{lrrrr}
    \toprule
    \textbf{Arm} & \textbf{Semantic} & \textbf{Logic} & \textbf{Expression} & \textbf{Weighted} \\
    \midrule
    A0 & 0.628 & 0.620 & 0.706 & 0.640 \\
    A1 & 0.762 & 0.744 & 0.764 & 0.755 \\
    A2 & 0.388 & 0.398 & 0.806 & 0.475 \\
    A3 & 0.782 & 0.776 & 0.788 & 0.781 \\
    A4 & 0.829 & 0.841 & 0.840 & 0.836 \\
    A5 & 0.849 & 0.845 & 0.870 & 0.852 \\
    \textbf{A6} & \textbf{0.915} & \textbf{0.930} & \textbf{0.918} & \textbf{0.921} \\
    \bottomrule
  \end{tabular}
\end{table}

%% file: sections/12_conclusion.tex
\section{Conclusion}

\modelname treats mobile Query prediction as a connected quality--efficiency problem rather than a sequence of isolated optimizations. The system first turns noisy cross-surface activity into a compact, evidence-preserving trajectory. It then builds a recommendation-native foundation, aligns that foundation with a strict Query contract and structured reasoning, and uses constrained reinforcement learning to remove computation that does not improve the decision. Quantization, structured compression, distillation, latent thinking, and device--cloud routing carry the resulting capability into a practical serving path.

The empirical evidence shows that direct generation is appropriate when behavior already contains a decisive signal. Ambiguous and post-purchase intents benefit from reasoning, but most of that benefit comes from selecting and relating the right evidence rather than producing a long explanation. The RL ablation makes the optimization requirement concrete: quality-only learning improves the Query while lengthening CoT, and ungated length pressure eventually damages grounding. Quality gating repairs that incentive, dynamic budgets remove unnecessary computation, and multiplicative rank-safe shaping reaches $78.6\%$ Query quality with a $14$-token median CoT and a $1.6\%$ hard-failure rate. The generated-Query path also reaches semantic regions that are less redundant with established recall channels, giving the mechanism a plausible route to downstream value.

The broader lesson is that efficient reasoning should be defined by sufficiency, not brevity. A mobile model should perform only the computation that the request requires while preserving the evidence that makes the final Query grounded and useful. This unifies domain learning, adaptive reasoning, and deployment compression around one measurable quality--cost frontier.

%% file: sections/appendix_contributors.tex
\clearpage
\section*{Appendix}
\section{Contributors}
\label{app:contributors}

\begingroup
\setlength{\parindent}{0pt}
\noindent
\begin{minipage}[t]{0.46\textwidth}
\vspace{0pt}
{\color{reccore}\bfseries Core Contributors\par}
{\color{reccore}
Lingqing Zhang\textsuperscript{\dag}\par
Bin Zhang\par
Weipeng Huang\par
}

\vspace{1.1em}
{\color{recnavy}\bfseries Contributors\par}
{\color{recnavy}
Chengfei Lv\par
Chengyu Lai\par
Chuxin Chen\par
Dimin Wang\par
Han Zhu\par
Hongtao Cheng\par
Jialin Zhu\par
Jian Wang\par
Jiuning Lin\par
Junqing Wu\par
}
\end{minipage}
\hfill
\begin{minipage}[t]{0.46\textwidth}
\vspace{0pt}
{\color{recnavy}
Li Chen\par
Qichao Ma\par
Ruiquan Lan\textsuperscript{\dag}\par
Shuai Zhong\textsuperscript{\dag}\par
Tao Wang\par
Xiaodong Zhu\textsuperscript{\dag}\par
Yinjiang Cai\par
Yinnan Song\par
Yipeng Yu\par
Yuan Liu\par
Yuning Jiang\par
Zhaode Wang\par
Zhibo Xiao\par
Zhixin Ma\par
Zihong Huang\par
}
\end{minipage}

\vspace{0.8em}
\noindent Core contributors are listed in contribution order. Contributors are listed alphabetically by first name.

\vspace{0.6em}
\noindent\textsuperscript{\dag} Work done during a summer internship at Taobao \& Tmall Group of Alibaba.
\endgroup

\clearpage

%% file: sections/appendix_prompt_examples.tex
\section{Prompt Examples}
\label{app:prompt-examples}

This appendix presents representative English prompt contracts used by the reasoning-data, relation-evaluation, and Query-generation stages. Each example contains the role instruction, behavioral evidence, and output schema needed to understand the method.

\subsection{Reasoning-Need Annotation and Structured Summary}

Under this contract, the model decides whether a target Query requires an explicit reasoning summary. When reasoning is necessary, the contract requires the output to decompose the evidence and decision into five auditable stages and the final Query to match the supplied target exactly.

\begin{systemprompt}[System]
\promptinput{prompts/reasoning_need_system.txt}
\end{systemprompt}

\begin{userprompt}[User]
\promptinput{prompts/reasoning_need_user.txt}
\end{userprompt}

\subsection{Post-Purchase Complementarity Judge}

The relation judge estimates whether a candidate Query represents a plausible complement to an observed product, coordinates stylistically with that product, or completes its usage scenario. Its structured score is used as one component of offline quality assessment rather than as a standalone optimization target.

\begin{systemprompt}[System]
\promptinput{prompts/relation_judge_system.txt}
\end{systemprompt}

\begin{userprompt}[User]
\promptinput{prompts/relation_judge_user.txt}
\end{userprompt}

\subsection{Explicit-CoT Query Generation}

The supervised generation contract separates a concise reasoning region from a single final Query. The final region permits exactly one search expression and excludes explanations or alternative candidates.

\begin{systemprompt}[System]
\promptinput{prompts/query_generation_system.txt}
\end{systemprompt}

\begin{userprompt}[User]
\promptinput{prompts/query_generation_user.txt}
\end{userprompt}